\documentclass[usegraphicx,usenatbib]{mn2e}
\input{epsf}
\usepackage[usenames]{color}
\usepackage{ amssymb }

\def\lsim{ \lower .75ex \hbox{$\sim$} \llap{\raise .27ex \hbox{$<$}} }
\def\gsim{ \lower .75ex \hbox{$\sim$} \llap{\raise .27ex \hbox{$>$}} }

\begin{document}
	
\title[LBGs UV colours and attenuation]{The UV colours
  and dust attenuation of Lyman-break galaxies.}
 
\author[]{
\parbox[t]{\textwidth}{
\vspace{-1.0cm}
V.\,Gonzalez-Perez$^{1,2}$,
C.\,G.\,Lacey$^{1}$,
C.\,M.\,Baugh$^{1}$,
C.\,S.\,Frenk$^{1}$.
S.\,M.\,Wilkins$^{3}$
}
\\
$^{1}$Institute for Computational Cosmology, Department of Physics, 
University of Durham, South Road, Durham, DH1 3LE, U.K.
\\
$^{2}$Centre de Physique des Particules de Marseille, Aix-Marseille
Universit\'e,  CNRS/IN2P3, F-13288 Marseille cedex 9, France.
\\
$^{3}$University of Oxford, Department of Physics, Denys Wilkinson
Building, Keble Road, OX1 3RH, U.K.
}
 
\maketitle
\begin{abstract}
Using {\sc galform}, a semi-analytical model of galaxy formation in the
$\Lambda$ cold dark matter cosmology, we
study the rest-frame ultraviolet (UV) colours of Lyman-break galaxies (LBGs) in the redshift range
$2.5\le z\le 10$. As the impact of dust on UV luminosity can be
dramatic, our model includes a
self-consistent computation of dust attenuation based on a radiative
transfer model. We find that intrinsically brighter galaxies
  suffer stronger dust attenuation than fainter ones, though the
  relation has a large scatter. The model predicts galaxies
  with UV colours consistent with the colour selection regions
  designed to select LBGs in observational surveys.  We find that the drop-out technique that
  selects LBGs based on two rest-frame UV colours is robust and
  effective, selecting more than 70 per cent of UV bright galaxies at a
  given redshift. We investigate the impact on the
  predicted UV colours of varying selected
  model parameters. We find that the
  UV colours are most sensitive to the modelling of dust attenuation and, in particular, to the
  extinction curve used in the radiative transfer calculation. If we assume a Milky Way dust extinction curve, the predicted
  UV continuum slopes are, in general, bluer than
  observed. However, we find that the opposite is true when using the Small
  Magellanic Cloud dust extinction curve. This demonstrates the strong dependence of UV
  colours on dust properties and highlights the inadequacy of using the UV
  continuum slope as a
  tracer of dust attenuation without any further
  knowledge of the galaxy inclination or dust characteristics in high redshift galaxies. 
\end{abstract}

\begin{keywords}
galaxies: evolution, galaxies: formation, galaxies: high-redshift.
\end{keywords}

\section{Introduction}

The recent availability of near infrared (IR) imaging on the Wide Field
Camera 3 (WFC3) on board the Hubble Space Telescope (HST), extending the
wavelength coverage to span 2000\AA\    to 17000\AA,  has
dramatically increased the number of candidate high redshift  galaxies
with $z>2.5$ \citep[e.g.][]{hathi10,oesch10,sw11a,sw11,mclure11,bou12}. The
galaxy redshift ``record'' has been tentatively pushed up to $z=10$, with 3
candidates awaiting spectroscopic confirmation \citep{yan10,bou10}.

Lyman-break galaxies are selected by targeting spectral features of
star-forming galaxies at $z\ge 2.5$, in particular the Lyman break, a
combination of both the Lyman-limit break (at 912\AA) due to neutral hydrogen in
each galaxy and the Lyman-$\alpha$-break (at 1216\AA) due to absorption by intervening neutral gas
clouds,  and the UV continuum slope.  \citet{steidel96} proved the usefulness of this
technique by spectroscopically confirming drop-out candidates isolated
photometrically at
$z\sim 3$. Since then, this approach has been adapted to find higher
redshift galaxies using both ground and space based telescopes \citep{steidel99,bou03,shimasaku05,yoshida06}.

At these high redshifts, we are witnessing the
rapid evolution of very young galaxies, when the Universe was less
than a quarter of its current age. Studies of galaxies selected using the drop-out technique have been
fundamental in unveiling the global star formation history of
the Universe \citep{madau96,steidel99}. Moreover, observations
at redshifts above $z=6$ could help us to understand the end of the
reionisation of the intergalactic medium
\citep[e.g.][]{robertson10}. However, the interpretation of the observational data is likely to be significantly affected by
the attenuation of starlight due to dust \citep[e.g.][]{chapman05}. 

Currently, there is some
controversy over the level of dust attenuation inferred in LBGs. \citet{mclure11} and \citet{finkelstein12} estimated the far UV
attenuation using spectral energy distribution (SED) fitting for
LBGs brighter than $M_{AB}(1500$\AA$)-5{\rm log}h=-17$. Both studies used template spectra including dust
attenuation following the Calzetti law. \citeauthor{finkelstein12} inferred an attenuation at
1500 \AA, $A_{1500}$, of $\sim 1.3$ magnitudes at $z\sim 4$ and $A_{1500}<0.25$ at
$z\sim 7$. In contrast, for galaxies with
$\langle z \rangle=6.5$, \citeauthor{mclure11} found  $\langle
A_{1500}\rangle \sim 0.4$, a value above the upper limit found by \citeauthor{finkelstein12} In many observational studies the dust attenuation is
inferred from the UV continuum slope estimated from a single
colour.  \citet{bou10} measured an average UV continuum slope of -3 for
galaxies at $z=7$. However, this value was measured to be $\sim-2.2$ when more data were collected by the HST WFC3  \citep{bou12}. This illustrates how the scarcity of high redshift data
can bias the estimation of dust attenuation. 

Dust can absorb and scatter photons either into or out of the line of sight to
the observer. The outcome of these
processes depends strongly on the geometry of dust and stars in galaxies and a
full radiative transfer model is required for a realistic calculation of
the effect of dust on star light
\citep{fontanot09,fontanot11}. Observationally, the attenuation by dust of the UV light emitted by LBGs is usually estimated
from the observed UV continuum slope
\citep{meurer99}. \citet{calzetti00} proposed a
universal shape for the dependence of the attenuation on
wavelength, such that the UV light is the most affected by dust. This so called ``Calzetti
law'' is based on observations of 39 nearby UV-bright starburst galaxies. However, the Calzetti law
provides a poor fit to the UV colours of larger samples of nearby
galaxies that have a range of star formation rates
\citep{conroy10,buat11,wild11}. Other observational studies also find
  that, at least for part of the UV-selected galaxies at $1<z<4$, the attenuation
is not well described by the Calzetti law \citep{noll09,buatbump,buat12,reddy12,lee12}.

In this paper we use a semi-analytical approach to model the formation
and evolution
of galaxies with a
realistic treatment of dust. In
particular, we study the rest-frame UV colours of LBGs within a
cold dark matter universe using {\sc galform}, the semi-analytical model
developed initially by \citet{cole00}. 

Previous theoretical studies of LBGs include both semi-analytical \citep{baugh98,guo09,lofaro09,lacey11,somerville12} and
gas-dynamical simulations of galaxy formation
\citep{nagamine02,finlator06,jaime10,cen11,jaacks12,pratika12}. This latter
type of simulation has the drawback that, on the whole, they do
not predict a present-day galaxy population consistent with
observations, unlike the semi-analytical models. Furthermore, most of
these models treat the attenuation by dust by using either the
Calzetti law or the slab model. \citet{guo09}, \citet{lofaro09} and
\citet{somerville12} investigated the effects on the
inferred luminosity functions and other properties of applying
observational LBG colour selections to model galaxies. \citeauthor{guo09} used
a phenomenological model for dust attenuation, while  \citeauthor{lofaro09} used a
physical model similar to that in the present paper, but restricted
their analysis to LBGs at $4\le z\le 6$. \citeauthor{somerville12}
studied galaxies at $0\le z \le 5$, calculating the dust attenuation
using a slab model with two dust components, diffuse and molecular
clouds. They found that in order to reproduce the observations, they
needed to let the effective optical depth normalisation evolve with redshift.

In this study, we use the version of {\sc galform} developed by
\citet{baugh05} model, but with the theoretical treatment of dust attenuation described in
\citet{lacey11}, in place of the {\sc grasil} \citep{grasil} model, which was
used in the original paper. {\sc grasil} is a
spectrophotometric code which is quite expensive
computationally and, thus, difficult to run for large samples of galaxies. The parameters of the
\citet{baugh05} model were set in order to
match observations of sub-mm selected galaxies, with a median redshift
$z\sim 2$, and the LBG luminosity function at $z\sim 3$ within a
single framework, at the same time as reproducing the properties of
local galaxies. \citet{lacey11} found that the
\citeauthor{baugh05} model predicts UV luminosity functions that agree with observations up to
$z=10$, the highest redshift for which measurements are currently
available. This agreement is notable as most of these observations were not
available at the time the model was proposed. A typical
attenuation in the UV of about two magnitudes was
predicted for bright galaxies, posing the questions of whether or not a model with such
strong attenuation requires galaxies with unrealistic dust contents, and if it
could produce UV colours consistent with those observed. 

Here, we show that the Baugh et al. model produces galaxies with the
right colours to be selected as LBGs over the redshift interval $3\leq
z\leq 10$. We also investigate the attenuation by dust in the model. Intrinsically bright
galaxies are predicted to be heavily extincted in the
model. Nevertheless, the UV continuum slope inferred from the attenuated photometry
of model galaxies is {\it not} necessarily redder than observed. In fact, our
calculations show that the UV continuum slope is a poor indicator of dust
attenuation, as our results are extremely sensitive to the choice of extinction
curve used as the input to the attenuation of starlight calculations. For
our standard choice, a Milky Way (MW) extinction curve, galaxies get {\it
  bluer} when they are attenuated, which, as we explain in Section
\S\ref{sec:beta} is due to the
presence of a bump at 2175\AA\ in the extinction curve. The
discrepancy between the model predictions and the observations for the slope of the attenuated UV
continuum slope is smaller than the difference in the predictions
on replacing the MW extinction curve by an SMC-like extinction curve.

In section \S\ref{sec:model} we
summarise the main features of the \citet{baugh05} model. The drop-out
technique is introduced in section \S\ref{sec:dropout}. Section
\S\ref{sec:Mdust} presents the predicted dust masses and attenuation
in the UV for bright galaxies at different redshifts. In sections
\S\ref{sec:colours} and \S\ref{sec:beta} we explore the predicted UV
colours and UV continuum slope for galaxies with $z>2.5$. The
conclusions may be found in \S\ref{sec:conclusions}.

All the magnitudes quoted are on the AB system, unless stated
otherwise.

\section{Galaxy formation model}\label{sec:model}

We predict the UV luminosities and colours of galaxies in a $\Lambda$
Cold Dark Matter ($\Lambda$CDM) 
universe using the {\sc galform} semi-analytical galaxy formation 
model initially developed by \citet{cole00}. Semi-analytical models 
use simple, physically motivated equations to follow the 
fate of baryons in a universe in which structure grows hierarchically 
through gravitational instability \citep[see][for an overview of 
hierarchical galaxy formation models]{baugh06,benson10}. 

{\sc galform} models the main processes which shape the formation and
evolution of galaxies. These include: (i) the collapse and merging of
dark matter haloes; (ii) the shock-heating and radiative cooling of
gas inside dark matter haloes, leading to the formation of galaxy
discs; (iii) quiescent star formation in galaxy discs; (iv) feedback
from supernovae (SNe) and from photoionization of the intergalactic medium (IGM); (v) chemical enrichment of the stars and gas; (vi) galaxy
mergers driven by dynamical friction within common dark matter haloes,
leading to the formation of stellar spheroids, which also may trigger
bursts of star formation. The end product of the calculations is a
prediction for the number and properties of galaxies that reside
within dark matter haloes of different masses.

We focus our attention on the \citet{baugh05} model.  Some of
the key features of this model are (i) a time-scale for quiescent
star formation that varies simply as a power of the disc circular
velocity  \citep[see][for a study of
  different star formation laws in quiescent galaxies]{lagos10}, (ii) bursts of star formation are triggered only by galaxy
mergers, (iii) a \citet{kennicutt_imf} initial mass function (IMF) is adopted in quiescent star formation in galactic discs, while in starbursts a top-heavy IMF is assumed, with slope
$x=0$, (iv) the inclusion of SNe feedback with the possible
occurrence of superwinds  \citep[see][for a discussion of
  the effect that feedback has on the luminosity function of
  galaxies]{benson03}, and (v) feedback from the ionisation of the IGM is approximated by a simple model in which gas cooling is
completely suppressed in haloes with circular velocities less than
$30$km s$^{-1}$ at redshifts $z<10$ \citep{lacey11}. In sections
\S{\ref{intro:dust}} and \S{\ref{intro:IGM}}, we give a detailed
account of how the attenuation due to dust and the IGM are implemented in this model. The parameters of this model were 
fixed with reference to a subset of the available observations of galaxies, 
mostly at low redshift.  The \citeauthor{baugh05} model uses the canonical ($\Lambda$CDM) 
parameters: matter density, $\Omega_{0}=0.3$, cosmological constant, 
$\Lambda_{0} = 0.7$, baryon density, $\Omega_{b}=0.04$, a normalisation of 
density fluctuations given by $\sigma_{8}=0.93$ and a Hubble constant
today of $H_0=100h$ km$\,{\rm s}^{-1}$Mpc$^{-1}$, with $h=0.7$. The model employs merger trees generated using 
a Monte Carlo algorithm. Here we use the algorithm introduced by
\citet{parkinson08}. \citet{juansubmm} reported that using this
particular approach does not change any of the original model predictions. We refer the 
reader to \citet{baugh05},  \citet{lacey08} and \citet{lacey11} for a full description
of this model.

In addition to reproducing local galaxy data, the \citeauthor{baugh05}
model matches the number and redshift distribution 
of galaxies detected by their emission at sub-millimetre wavelengths 
\citep{baugh05,lacey08}, the rest-frame UV luminosity 
function of Lyman-break galaxies up to $z\sim 10$ \citep{lacey11} and the abundance and clustering of 
Lyman-$\alpha$ emitters \citep{orsi08}. \citet{milan11} used the
\citeauthor{baugh05} model to explore the emissivity of ionising photons at $z\ge
6$. \citet{juanlbgs} showed that the \citeauthor{baugh05} model
predicts that most present day galaxies with masses close
to that of the MW had a LBG progenitor at $3<z<4$.

No parameters have been tuned for the study presented here. We do vary
some of the parameters in the model in \S\ref{sec:colours} and \S\ref{sec:beta} to illustrate the sensitivity
of the model predictions to our choices. 

\subsection{Dust attenuation}\label{intro:dust}
Following the approach presented in
\citet{cole00} and \citet{lacey11,lacey12}, the attenuation of starlight by dust is modelled in a physically
self-consistent way, based on the results of a radiative transfer calculation for a realistic
geometry in which stars are distributed in a disk plus bulge. This
geometry is the same as that assumed in more detailed computations
carried out in the {\sc grasil} code
\citep{grasil}. An important feature of the method is that the
dust attenuation varies self-consistently with other galaxy properties
such as size, gas mass, and metallicity.

Dust can absorb photons and scatter them either into or out of the line of sight to
the observer. Here we refer to dust
``attenuation'' as the average effect that dust has on starlight
once the geometric configuration of dust and stars is taken into
account. Dust attenuation includes the effects of both light absorption
and scattering, which strongly depend on how dust is distributed with
respect to the stars in a galaxy. We can express the dust attenuation, a$_{\lambda}$, as
the ratio between the attenuated and unattenuated luminosity of a galaxy:
\begin{equation}
{\rm a}_{\lambda}=\frac{L_{\lambda}({\rm attenuated\, by\,
    dust})}{L_{\lambda}({\rm unattenuated\, by\, dust})}.
\end{equation}
The effective optical depth at a given
wavelength can be related directly to the dust attenuation at that wavelength:
$\tau_{{\rm eff,}\, \lambda}=-{\rm ln}({\rm a}_{\lambda})$. We can also
express the attenuation in terms of magnitudes, ${\rm A}_{\lambda}$,  and then $\tau_{{\rm eff,}\,
  \lambda}={\rm A}_{\lambda}/(2.5{\rm log} e)$\footnote{Throughout all
the paper we use the notation log to designate the logarithm in base
10.}.

Usually the term dust ``extinction'' refers to the attenuation of the
light from a point source placed behind a screen of dust. Thus, the
``extinction'' is independent of the geometry of the system.

In {\sc galform}, dust is assumed to be in two
components: dense molecular clouds embedded in a diffuse
component. Stars inside molecular clouds have their light attenuated both by the clouds and by the
diffuse medium, while the remaining stars are attenuated only by the
diffuse medium.

\subsubsection{Diffuse component}
The attenuation of starlight by the diffuse component of dust is
calculated using the radiative transfer models of
\citet{ferrara99}. These models assume that stars are distributed in
two components: an exponential disk and a bulge; and that dust is
smoothly distributed in the same plane as the stellar disk. Given an
extinction law, by default the MW one, the \citet{ferrara99}
models give dust attenuation factors separately for the disc and
spheroid light, as a function of wavelength, galaxy
inclination,
ratio of bulge to disc radial dust scalelength, $r_{e}/h_{R}$,
ratio of dust to stellar vertical scaleheights,
$h_{z,dust}/h_{z,stars}$, and $\tau_{V_0}$. Here  $\tau_{V_0}$ is the
extinction optical depth  in the V-band looking through the centre of a face-on
galaxy.  In {\sc galform}, the
diffuse component of dust is assumed to follow the distribution of
stars in a disk or in a starburst component. By default, the dust is assumed to have the same radial and
vertical scalelengths as the stars. The rest of the above properties are predicted by
{\sc galform}. 

In order to calculate the extinction
optical depth in the V-band, $\tau_{V_0}$, the extinction $A_{\rm V}$ along the line-of-sight is determined
from the neutral hydrogen column density, $N_{\rm H}$, and the cold
gas metallicity, $Z_{\rm cold}$:
\begin{equation}
A_{\rm V}=\Big(\frac{N_{\rm H}}{{\rm cm}^{-2}}\Big)\Big(\frac{Z_{\rm cold}}{Z_{\odot}}\Big)\Big(\frac{A_{\rm V}}{N_{\rm H}}\Big)_{\odot}.
\end{equation}
The above extinction is normalised in such a way that gas
with solar metallicity, Z$_{\odot}=0.02$, has the local interstellar
medium (ISM)
dust-to-gas ratio \citep[e.g.][]{guiderdoni87}.  The neutral hydrogen column density is related to
the projected central surface gas mass in the disk, $M_{\rm cold}/(2\pi h_R^2)$, and thus:
\begin{equation}
\tau _{V_0}=0.043\Big[\frac{M_{\rm cold}}{2\pi
      h_{R}^2}\frac{pc^2}{M_{\rm \odot}}\Big]\Big(\frac{Z_{\rm cold}}{0.02}\Big).
\end{equation}
The extinction optical depth, $\tau_{V_0}$, is obtained in the same way independently
of which extinction curve is being used afterwards for the calculation
of the attenuation. 

\begin{figure}
{\epsfxsize=8.5truecm
\epsfbox[93 370 344 650]{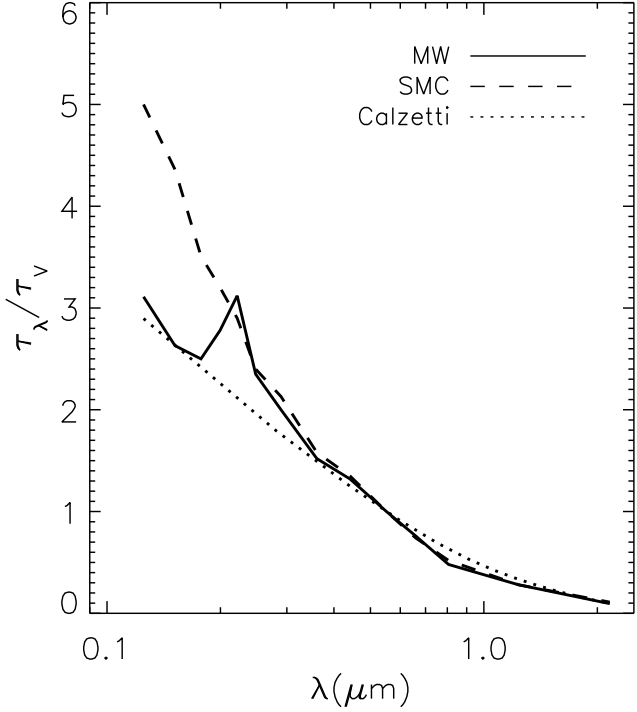}}
\caption
{The optical depth normalised by its value in the V-band as a function
  of wavelength. We plot the extinction curves from the MW
  (solid line) and SMC (dashed line) used
by \citet{ferrara99} to estimate the attenuation curves of
galaxies, and the attenuation curve proposed by
  \citet{calzetti00} to describe observations of nearby star
  forming galaxies (dotted line).}
\label{fig:att}
\end{figure}

Fig. \ref{fig:att} shows the normalised optical depth in the UV to near IR
wavelength range, for the extinction curves from the MW
and SMC used
by \citet{ferrara99} to estimate the attenuation curves of
galaxies. For comparison, Fig. \ref{fig:att} also displays the
attenuation curve empirically derived from observations of nearby
UV-bright, starburst galaxies by \citet{calzetti00}. For wavelengths
$\lambda \ge 0.35 \mu$m, the three curves are roughly consistent with each other. However, at smaller wavelengths the SMC
extinction curve displays much higher extinction than both the MW and
Calzetti curves. The MW extinction curve presents a bump at $\lambda \sim 2175$\AA\             that is not
seen in the other two curves. The presence of this bump is related
to the existence of small carbonaceous particles that could be
destroyed by the ambient UV radiation field \citep{fischera11}. Though less pronounced, such a bump has been
observed in the Large Magellanic Cloud \citep{gordon03} and in nearby
star-forming galaxies \citep[over 23000 galaxies at $z\sim
  0.07$,][]{wild11}. However, the bump at
$\lambda \sim 2175$\AA\  was not detected in the sample of 39
starburst galaxies analysed by \citet{calzetti94,calzetti00}. It is
worth noting here that \citet{vijh03}, using a radiative transfer
calculation, found the Calzetti attenuation law curve in the
particular case of assuming $\tau _{V_0}=1.5$ and a clumpy dusty shell with a SMC extinction curve.

\subsubsection{Molecular clouds}
In the model, stars are assumed to form in molecular clouds 
from which they will eventually leak out.  The attenuation of starlight by dust in molecular clouds is calculated
analytically, assuming new born stars to be a point source in the
centre of a
spherical cloud with uniform density. 

The attenuation by clouds depends both on the fraction of light produced
within the cloud and the extinction optical depth.  The light produced
within molecular clouds depends on the star formation and chemical
enrichment history, the IMF and the wavelength. The
attenuation by clouds mostly affects the UV light, which is mainly
produced by young stars with a short lifetime. Thus, the
fraction of light produced within a cloud is obtained by assuming
the recent star formation rate to be either constant in the case of quiescent
disks or to decay exponentially for bursts of star formation. 

The extinction optical depth is
calculated from the column density of gas through a cloud, assuming
the same extinction law as for the diffuse component of dust.

As expected, the attenuation calculated in this way depends on
wavelength, the metallicity of the gas and the
surface density of the clouds, which, in turn, depends on the cloud mass
and radius. For further details see Lacey et al. (2012, in prep).

The values of the parameters affecting the dust attenuation
calculation in molecular clouds are set to be the same as in the
\citeauthor{baugh05} model. The time for
stars to escape from their birth molecular clouds is set to t$_{\rm esc}=1$ Myr and the fraction of
dust in molecular clouds to f$_{\rm cloud}=0.25$. The value of f$_{\rm
    cloud}$ is consistent, within a factor of 2, with estimations
  based on observations of nearby galaxies \citep{granato00}. The
  value of t$_{\rm esc}$ was chosen in order to reproduce the
  luminosity function of observed LBGs at $z=3$. Moreover, the chosen value for t$_{\rm
    esc}$ is consistent with the escape time inferred from
  observations of stars in the MW \citep{tesc}. By default, the mass of clouds is set to be M$_{\rm cloud}=10^6
M_{\odot}$ and their radius to r$_{\rm cloud}=$16 pc, to match the choices made in
{\sc grasil} \citep{grasil}, that, in turn, were made to match
observations of giant molecular clouds in the MW and to fit the
spectral energy distributions (SED) of nearby galaxies. With these parameters the far-UV light
of stars within molecular clouds is almost completely extincted. Thus,
the net attenuation by clouds is insensitive to the exact value of the
surface density of a single cloud, which is proportional to $M_{\rm
  cloud}$/r$_{\rm cloud}^2$, and it will only depend weakly on the
assumed shape of the extinction curve used in the calculation of dust
attenuation in molecular clouds.

\subsubsection{Dust to metals ratios}

In {\sc galform}, a fixed fraction of the total
heavy elements in the gas component is assumed to be locked up in dust
grains. The dust to metal ratio is thus also fixed in the model. The model uses an instantaneous recycling approximation. This
approximation is reasonable as most metals are produced by Type II
SNe. Even at high redshifts, an appreciable difference in returned
mass fraction only persists for less than $10^{7.5}$yr \citep{nagashima05}. No attempt is
made to model in detail the formation and destruction of dust
grains. 

In principle, the dust to metal ratio depends on the dust production
mechanism, which in
turn depends directly on the star formation history of a galaxy. A
consensus on the dust production mechanism at different redshifts has
not yet been reached and thus, assuming a fixed dust to metal ratio
is a reasonable approximation. At low redshift, the bulk of the dust is thought to be generated by stars in the
asymptotic giant branch (AGB) phase. \citet{valiante09}
estimated that even at $z\sim 6.4$ between 50 and 80 per cent of the
total dust mass could be
generated by AGB stars. This percentage depends strongly on the assumed star
formation history.  AGB stars need to reach on average an
age of 400 Myr before they produce dust. Thus, at higher redshifts, $z\ge
6.5$, there is not enough time for typical AGB stars to generate the
dust that has been observed in some systems
\citep{dwek07,mm10}. Although SNe can generate large amounts of dust
they also destroy it in shocks \citep{bianchi07}. The most
plausible explanation for the observed dust at $z\ge 6.5$ appears to be the dust
growth in the ISM generated within dense molecular clouds
\citep{mm10,dwek11,valiante11}.

\begin{figure*}
\includegraphics[width=0.34\textwidth]{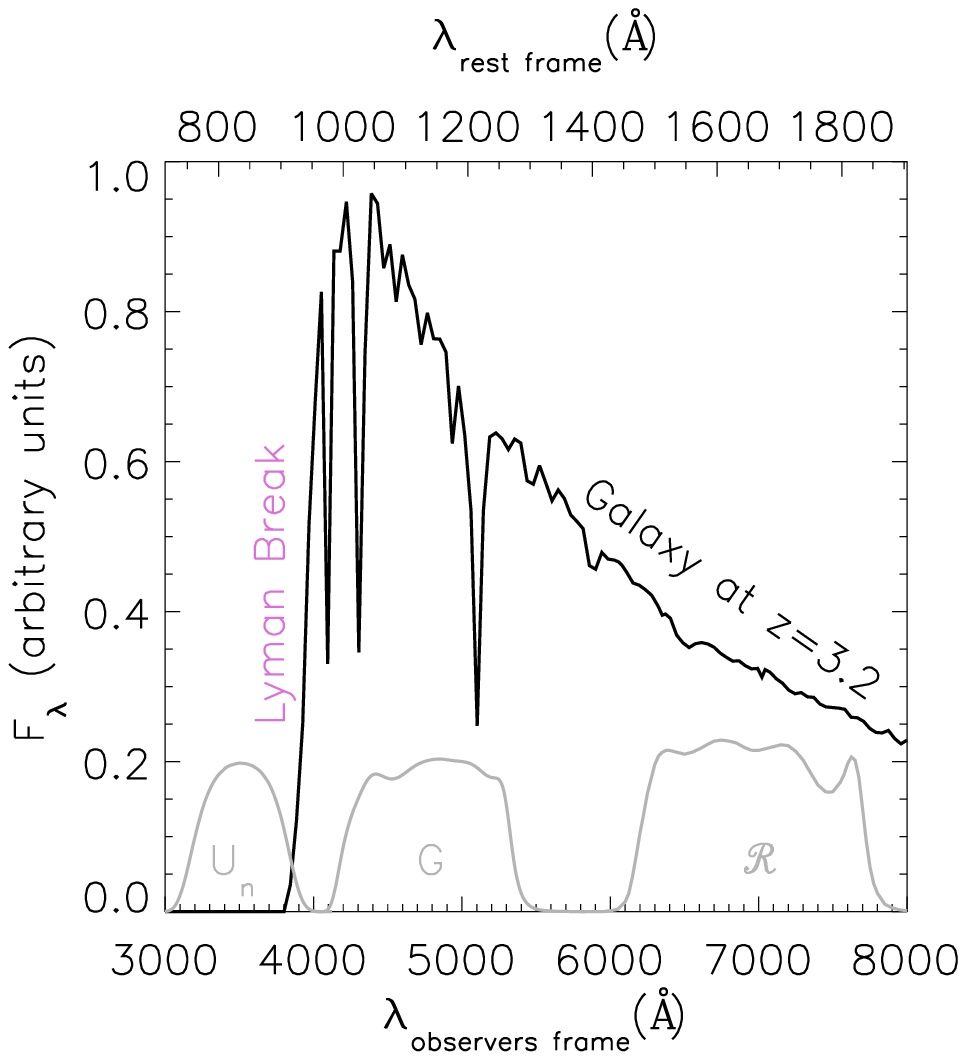}
\hspace{-0.5cm}\includegraphics[width=0.34\textwidth]{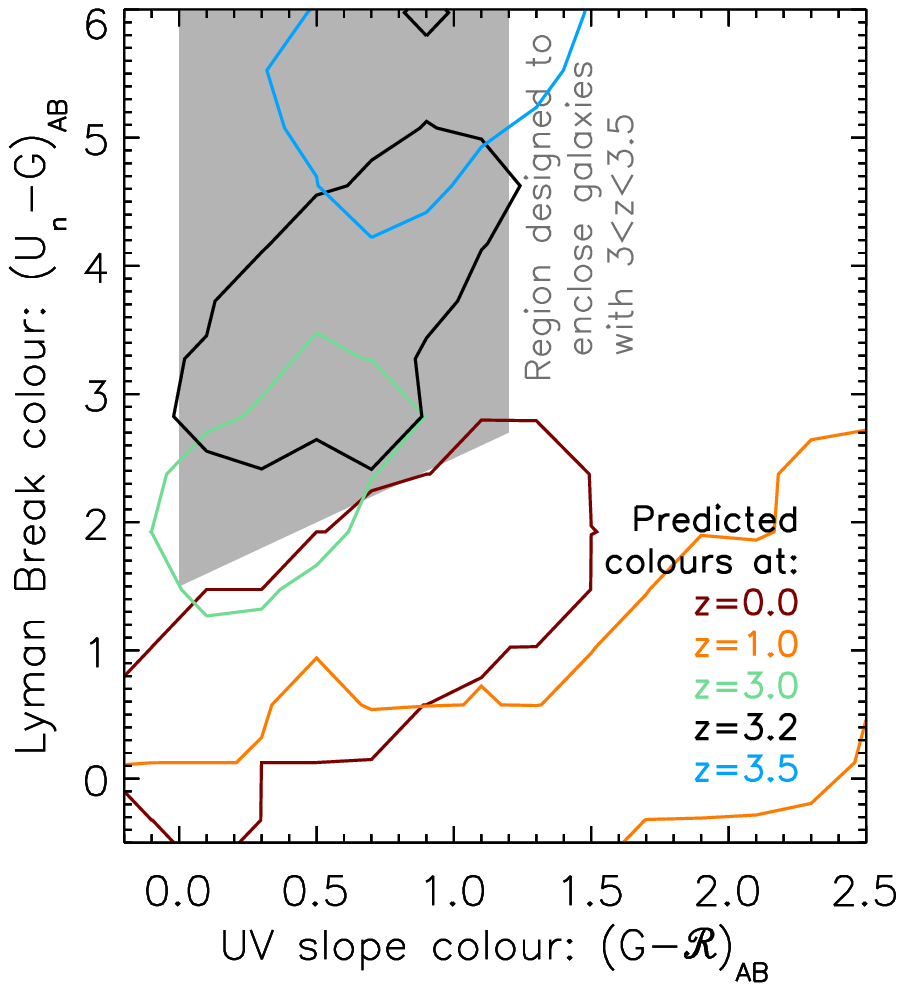}
\includegraphics[width=0.33\textwidth]{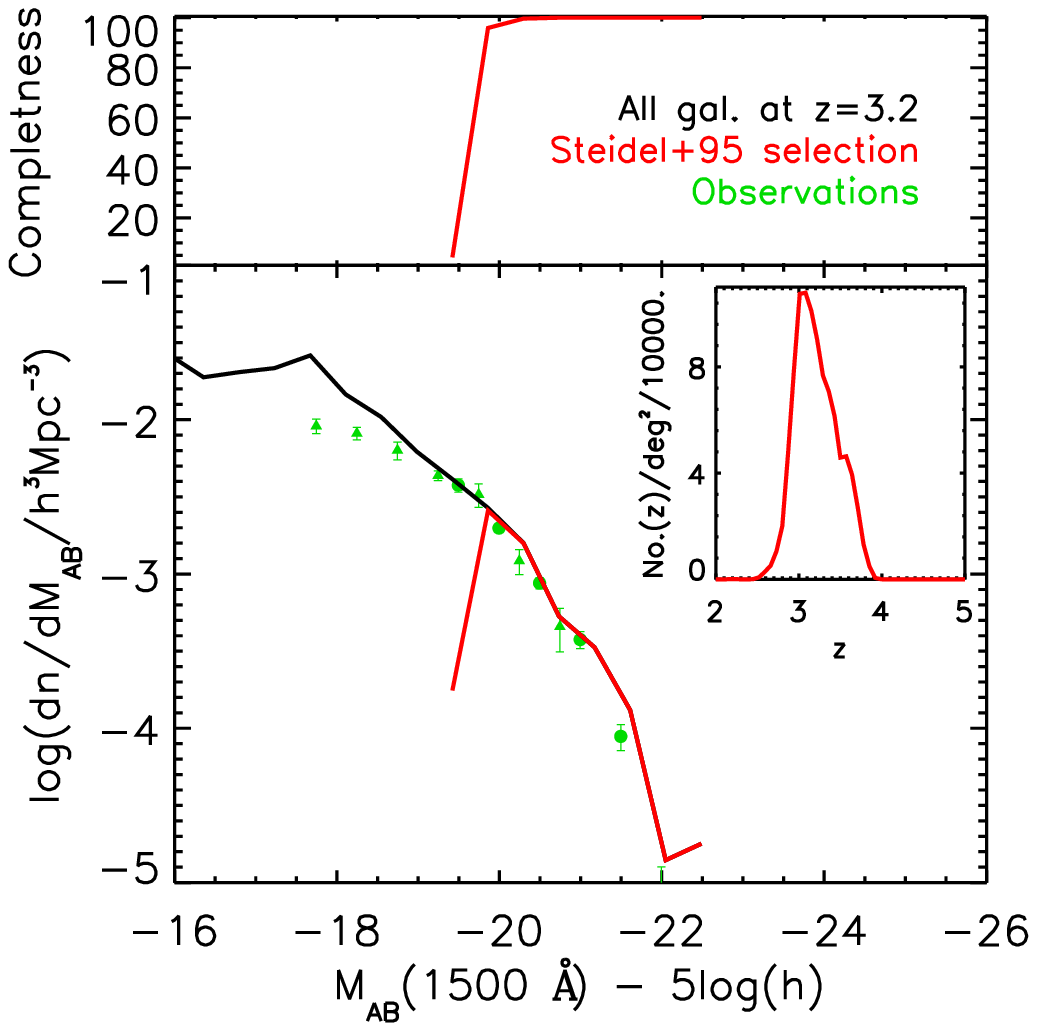}
\caption{An illustration of the drop-out technique used to select
  galaxies at $3<z<3.5$.
{\it Left panel:} The intrinsic spectrum of a star-forming galaxy at
  $z=3.2$, computed with the {\sc pegase.2} population synthesis model
  \citep{pegase} (black line, no attenuation is included) plotted together with the response
  functions of the U$_n$, G and $\mathcal{R}$ filters (grey
  lines). {\it Central panel:} The  (U$_n$-G) versus (G-$\mathcal{R}$)
  colour-colour plot proposed by \citet{steidel95} for selecting
  galaxies at $3<z<3.5$ (grey region) with the predicted density
  contours for galaxies at different redshifts, as indicated in the
  legend. The density contours enclose 80 per cent of the
  galaxies brighter than $\mathcal{R}<25.5$ and
  $G<27.29$ at each redshift. {\it Right panel:} The predicted
  completeness (top panel) and luminosity function (main panel)  at $z=3.2$
  of all galaxies (black line) and  galaxies with $\mathcal{R}<25.5$ and
  $G<27.29$ and the colour cuts shown in the central panel as a
  grey region (red lines). The redshift
  distribution of this last subsample is shown in the inset. For comparison, we
  show as green symbols the observed luminosity function from \citet{sawicki06} (triangles, 1700\AA), and \citet{reddy09}
  (circles, 1700\AA).
}
\label{fig:dropout}
\end{figure*}

\subsection{IGM attenuation}\label{intro:IGM}

The neutral hydrogen in the IGM absorbs and scatters photons from both the Lyman
series (bluewards of 1216\AA) 
and the Lyman continuum\footnote{The absorption due to the Helium present in the
IGM is, in general, negligible compared with that by neutral
hydrogen at the wavelengths of interest here.}. Within {\sc galform} this effect can be implemented
following the attenuation prescriptions of either \citet{ma} or
\citet{me}. The predicted
attenuation at 1000\AA\   due to the IGM increases
from about 0.5 magnitudes at $z\sim 4$ to $2$ magnitudes at $z\sim
6$, with either prescription.

\citet{ma} used an empirical approach to model the IGM attenuation,
while \citeauthor{me} based part of the prescription on the results of
a $\Lambda$CDM simulation with similar cosmological
parameters to those used in the B05 model.  Numerical simulations have
proved to be successful at reproducing different observational properties of absorbing
systems.

The effective optical depth through the IGM depends on the number of
absorbers, basically neutral hydrogen clouds, along the
line of sight. Each absorber is characterised by its redshift, its
neutral hydrogen column density, $N_{\rm HI}$, and its Doppler parameter, which is related to
the temperature and kinematics of the cloud. The absorption of the
Lyman continuum is mainly caused by Lyman limit systems (LLSs), with $N_{\rm HI}\sim
10^{17}$cm$^{-2}$ and scattering by the Lyman $\alpha$ forest (LAF), optically thin systems
at the Lyman edge, with  $N_{\rm HI}\sim
10^{13}$cm$^{-2}$. 

\citeauthor{ma} and \citeauthor{me} assumed different evolution with
redshift for the number of LAF and LLSs. While \citeauthor{ma} considered that the LAF followed a Poisson
distribution, \citeauthor{me} obtained the LAF distribution from a
N-body cosmological simulation.

An empirical evolution
of LLSs with redshift is assumed by both \citeauthor{ma} and \citeauthor{me}. However, only the latter
is consistent with current observations in the redshift range $0.3\lesssim
z \lesssim 4$ \citep[see Fig. 3
  in][]{inoue08}. 

Another important difference between the \citeauthor{ma} and \citeauthor{me}
modelling of the intergalactic attenuation is that in the former a fixed Doppler
parameter is assumed, while \citeauthor{me} varies this value based on
simulations, a
change supported by several observational studies
\citep[e.g.][]{kim97}. Moreover, the \citeauthor{ma} model
almost always predicts a smaller transmission than empirical models
based on the most recent observations \citep[see Fig. 6 in][]{inoue08}.

In light of the above discussion, the results presented in this paper are calculated by
default using the \citet{me} equations.  


\section{Lyman-break selection for galaxies at high redshifts}\label{sec:dropout}

Observationally, the most extensively used method for selecting star-forming galaxies at
high redshifts is the Lyman drop-out
technique \citep[for details on other methods see the review by][]{dunlopreview}. This technique uses near-UV, optical and near-IR filters
(depending on the target redshift) to sample the redshifted
Lyman-break, thereby selecting star-forming galaxies, the
so called Lyman Break Galaxies (LBGs). This method has been commonly
applied to select galaxies with $z>2.5$, since their Lyman-break is
shifted beyond the observed frame UV range that is difficult to observe with
ground-based telescopes. 

The drop-out technique typically uses 2 colours, one from
filters bracketing the Lyman-break and the second quantifying
the UV continuum slope. The rest-frame UV emission, $300\leq\lambda$(\AA)$\leq3000$, is largely due to the presence of young and
massive OB stars in galaxies. Thus, the rest-frame UV emission of
galaxies that are not actively forming stars is, in general, negligible in
comparison with that in star-forming galaxies. The Lyman-break is the
result of both the Lyman-limit break (at 912\AA) due to neutral
hydrogen in the atmospheres of
massive stars and the Lyman-$\alpha$-break (at 1216\AA) due to absorption by intervening neutral gas
clouds (e.g. the Lyman-$\alpha$ forest).

The drop-out technique was first proposed by
\citet{steidel92}, using the U$_n$, G and $\mathcal{R}$ filters
shown in Fig. \ref{fig:dropout}. The grey area in the central panel of
Fig. \ref{fig:dropout} shows the selection criteria used by
\citet{steidel95} to select galaxies at $3<z<3.5$. The left panel in Fig. \ref{fig:dropout} shows the
synthetic spectrum of a star-forming galaxy at $z=3.2$, computed with
the {\sc pegase.2} stellar population synthesis model \citep{pegase}
without including the attenuation by either dust or the IGM. This
spectrum shows a very pronounced break at 912\AA\ in the rest
frame, the Lyman-limit break. Bluewards of this
rest-frame wavelength, the galaxy is practically not emitting any
light. Thus, by measuring the (U$_n$-G) colour of a galaxy it is
possible to select candidates at $3<z<3.5$, since in this redshift
range the U$_n$ band is bluewards of the Lyman-break and the G band redwards. The (G-$\mathcal{R}$)
colour samples the rest-frame UV continuum slope and is used in order to minimise the number of lower redshift
interlopers.

The central panel of Fig. \ref{fig:dropout} shows, using density contours, the location in the colour-colour plane
of the model galaxies at different redshifts. It
is clear from this plot that a selection based only on the colour
sampling the Lyman-break (i.e. (U$_n$-G) in this example) will contain a substantial fraction of
galaxies at $z\sim 0$. Old and passively evolving galaxies at $z\sim 0$ present a deep enough break
at 4000\AA\ to be mistaken for the Lyman-break of star-forming
galaxies at higher redshifts. However, the slopes of the spectral
continuum redwards of these breaks are very different. The rest-frame
optical continuum of old galaxies at $z\sim 0$ is quite red, while the
rest-frame intrinsic UV continuum of star-forming galaxies is bluer, as
shown in the left panel of Fig. \ref{fig:dropout}. Thus, a cut on
(U$_n$-G) varying with (G-$\mathcal{R}$) can remove most of the low
redshift interlopers. However, the
presence of dust in star-forming galaxies at high redshift can redden
the (G-$\mathcal{R}$) colour, resulting in an overlap with the colours of lower redshift
galaxies. Thus, these high redshift dusty galaxies would not be
classified as LBGs.

The right panel of Fig. \ref{fig:dropout} shows the luminosity function of model galaxies selected using
\citeauthor{steidel95} magnitude and colour cut criteria, compared
with that for all model galaxies at $z=3.2$. This figure shows that the luminosity function of galaxies selected
following \citeauthor{steidel95} criterion is complete above the rest-frame UV
absolute magnitude that corresponds to the R-band magnitude limit
of the observations. Note that the observational data in the right
panel of Fig. \ref{fig:dropout} goes deeper than the original work from \citeauthor{steidel95}

The inset in the right panel in Fig. \ref{fig:dropout} shows the
predicted redshift distribution of galaxies with $\mathcal{R}<25.5$
and $G<27.29$ and colours within the
region proposed by \citeauthor{steidel95} to select galaxies at
$3\leq z \leq 3.5$. The model galaxies selected in this
way have redshifts in the range $3<z<4$, in agreement with
the selection proposed by \citeauthor{steidel95} 

\section{The predicted dust attenuation}\label{sec:Mdust}

The presence of even small amounts of dust in high redshift galaxies can have
a strong effect on the observed UV light. With the
same model as the one used here, \citet{lacey11} predicted that dust attenuation in the
far-UV ($\sim1500$\AA) is typically 2 magnitudes for galaxies
dominating the UV luminosity density, even
out to redshift 10. 

Below we present the predicted median dust masses, optical depths and
attenuations for galaxies at $3\leq z\leq 10$.

\subsection{The predicted dust masses}
\begin{figure}
{\epsfxsize=8.5truecm
\epsfbox[39 6 307 300]{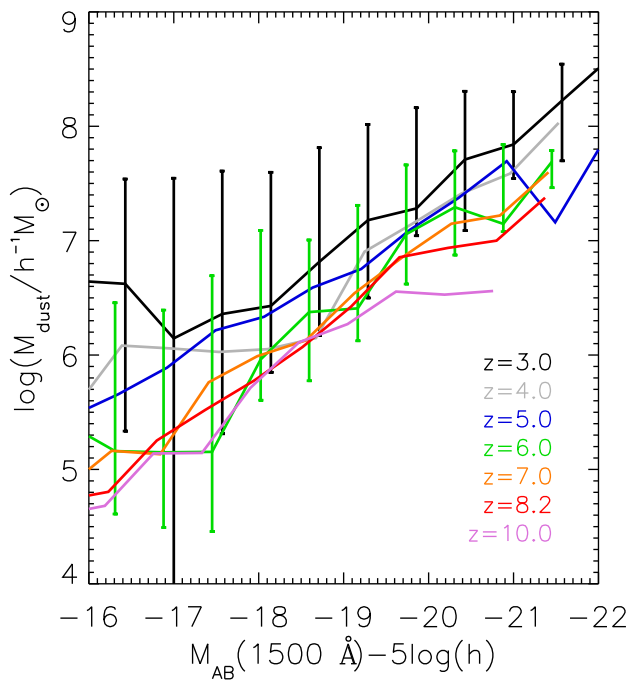}}
\caption
{Predicted median dust mass as a function of the attenuated absolute
  rest-frame UV
  magnitude ($\sim 1500$\AA) for galaxies at different redshifts, as indicated in the
  legend. For clarity, the error bars presenting the 10 and 90 percentiles are only shown for two redshifts.}
\label{fig:mdust}
\end{figure}

Following \citet{granato00}, who used a physical dust grain
model, we assume that the dust to gas ratio in
clouds and the diffuse ISM is proportional to the gas metallicity,
$Z_{\rm cold}$, with a value of $1/110$ for $Z=Z_{\odot}=0.02$. Thus, the
total dust mass in a galaxy, $M_{\rm dust}$, can be calculated as follows:
\begin{equation}\label{eq:mdust}
M_{\rm dust}=\frac{1}{110}X_{\rm H}M_{\rm cold}\Big(\frac{Z_{\rm cold}}{0.02}\Big),
\end{equation}
where $M_{\rm cold}$ is the
total cold gas mass in the galaxy and $X_{\rm H}=0.74$ is the mean hydrogen mass fraction in the Universe.

Fig. \ref{fig:mdust} shows the predicted median dust mass (Eq.\ref{eq:mdust}) as a
function of absolute rest-frame UV magnitude for galaxies at redshifts from 3 to
10. There is a clear trend for brighter galaxies to have higher dust
masses. This trend is a direct consequence of the predicted relation
between the far UV
magnitude and gas mass, combined with the fact that the predicted metallicity of cold
gas appears to be weakly dependant on luminosity \citep[see][]{lacey11}. It is
also clear from Fig. \ref{fig:mdust} that the model predicts a
decrease in dust content with increasing redshift. At a given
luminosity, we find a decrease in dust mass of a factor of $\sim$5 from $z=3$ to
$z=10$ for galaxies with
$-20\le M_{AB}$(1500\AA)$-5{\rm log}h\le -18$. 

Observations of galaxies in the wavelength range from the far IR to
the sub-mm have allowed direct estimates of the dust mass in some
galaxies above $z\sim 4$. Dust masses in excess of $\sim
10^8h^{-1}M_{\odot}$ have been found in galaxies with stellar masses estimated to be between $10^{10}$ to
$10^{11}$h$^{-1}M_{\odot}$ at redshifts
$4\leq z\leq 5$
and in QSOs up to $z\sim6.5$ \citep{mcmahon94,dunlop94,robson04,dwek07,mm10}. However, other galaxies at similar redshifts do not present any evidence
of containing dust \citep{zafar10}. For galaxies within the range of stellar masses $10^{10}$ to
$10^{11}$h$^{-1}M_{\odot}$, we find median dust masses around  $\sim
10^8h^{-1}M_{\odot}$, in agreement with those observations which
detect dust in high redshift galaxies.

\subsection{The predicted optical depth}
\begin{figure}
{\epsfxsize=8.5truecm
\epsfbox[30 4 304 300]{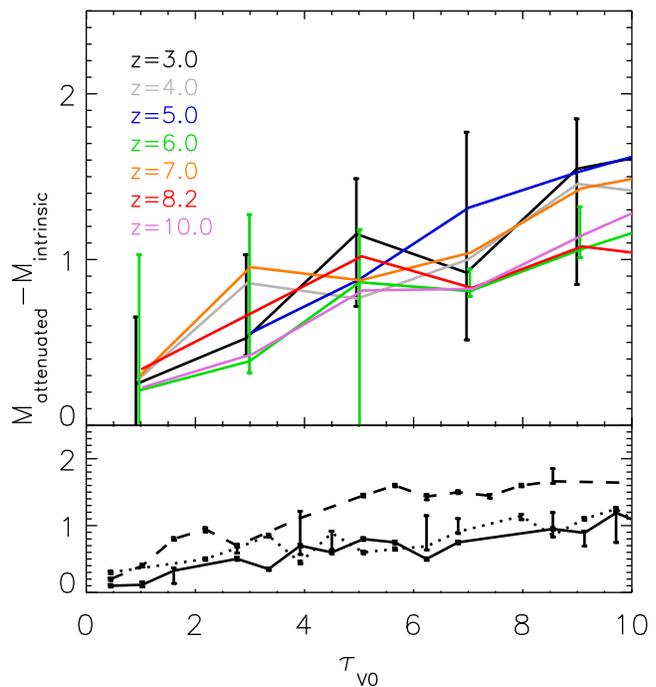}}
\caption
{ The predicted median difference between the attenuated and the
  intrinsic rest-frame UV magnitude as a function of the central
  extinction optical depth in the
  V-band,  $\tau_{V_0}$, for galaxies with M$_{AB}(1500$\AA$)-5$log$h<-17.8$, plotted at the
  redshifts indicated in the legend.  For clarity, the error bars
  representing the 10 and 90 percentiles are only shown for two
  redshifts. The {\it bottom panel} shows the attenuation at $z=3$ for
  galaxies binned in three inclination ranges: $15^o\le i <20^o$
  (solid line), $45^o\le i <50^o$ (dotted line, range bracketing the
  median value of the sample), $70^o\le i <75^o$
  (dashed line).}
\label{fig:tau}
\end{figure}

As described in \S\ref{intro:dust}, the dust attenuation of
starlight is directly related to the extinction central optical
depth in the V-band of a galaxy, $\tau_{V_0}$. Fig. \ref{fig:tau}
presents the median predicted attenuation in the UV as
a function of $\tau_{V_0}$ for
galaxies with M$_{AB}(1500$\AA$)-5{\rm log}h<-17.8$ \citep[corresponding to a tenth of the characteristic luminosity
  at $z=3$ as reported by][]{steidel99}. Despite the
scatter, there is a clear trend for
galaxies with higher extinction optical depths to present higher UV attenuations. This trend is found to be independent of redshift, in the studied range
$3\leq z \leq 10$. We find the median optical depth of galaxies brighter than
M$_{AB}(1500$\AA$)-5$log$h<-17.8$  to be between 2 and 6. These median values
depend strongly on the magnitude cut applied to select the galaxies. We
also find that brighter galaxies present higher extinction optical
depths, though this trend has a large
scatter. 

\subsection{The predicted median UV attenuation}
\begin{figure}
{\epsfxsize=8.5truecm
\epsfbox[28 0 297 258]{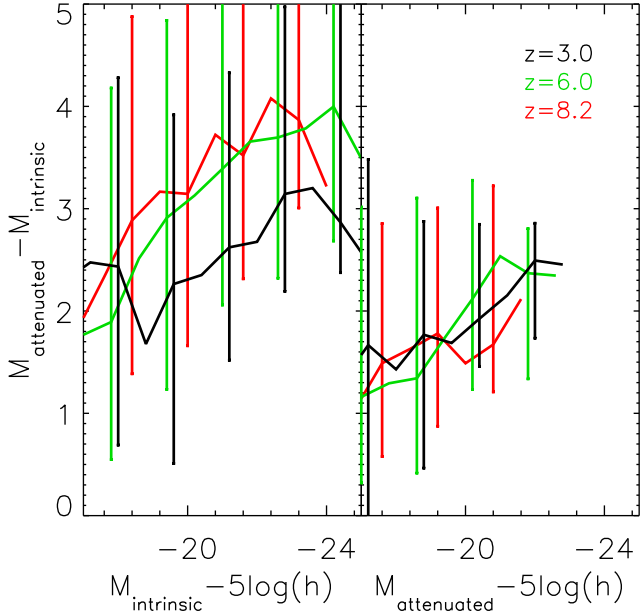}}
\caption
{The median attenuation in magnitudes at 1500\AA\ as a function of
  intrinsic (left panel) and attenuated (right panel) absolute UV (1500\AA) magnitude for model galaxies at
$z\sim 3$ (black), 6 (green) and 8.2 (red). The error bars show the 10
and 90 percentiles for the distribution.}
\label{fig:Dmagall}
\end{figure}
\begin{figure}
{\epsfxsize=8.5truecm
\epsfbox[28 0 297 258]{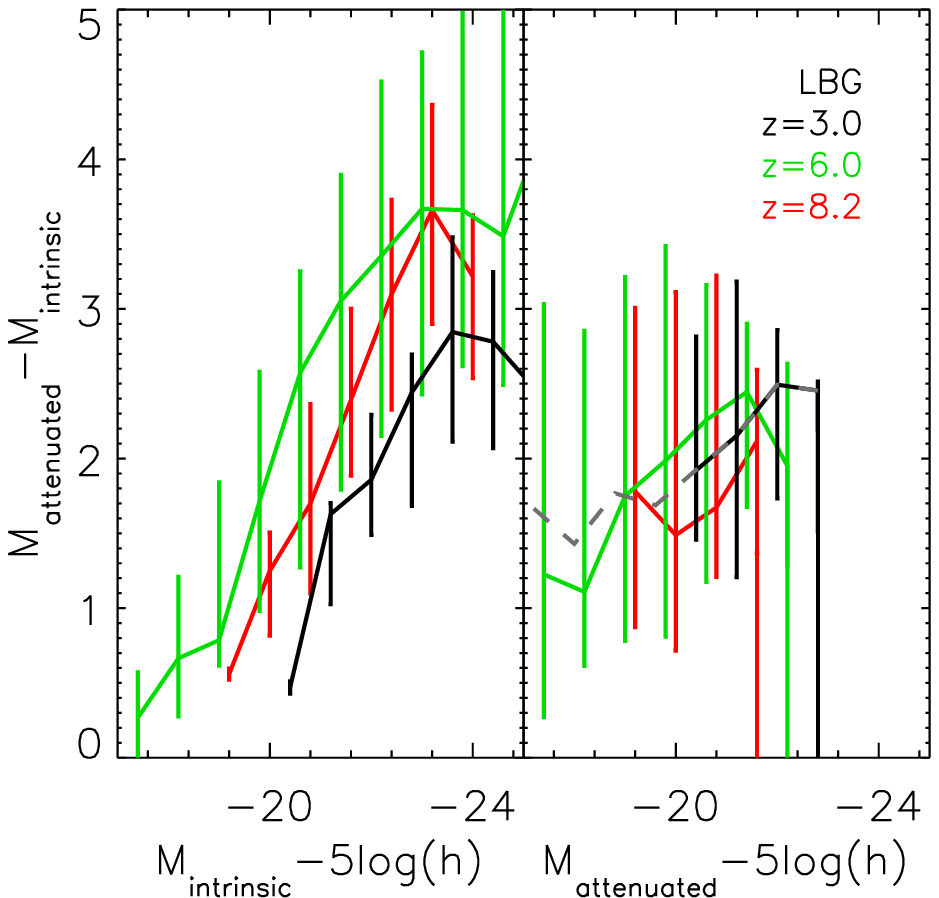}}
\caption
{Figure similar to Fig. \ref{fig:Dmagall} but for the subsample of
  model galaxies which pass various LBG colour and magnitude
  selections at $z\sim 3$ \citep[selection from][black]{steidel95}, 6
  \citep[selection from][green]{bou12} and 8.2 \citep[selection
  from][red]{lor11}. For comparison, the median attenuation for
  all galaxies at $z=3$, shown in Fig. \ref{fig:Dmagall}, is included
  here as a grey dashed line. The error bars show the 10
and 90 percentiles of the distribution.}
\label{fig:Dmaglbg}
\end{figure}

The attenuation of starlight depends not only on the extinction central optical depth but also on the
inclination of the galaxy, the distribution of the dust and the
fraction of dust in molecular clouds. The lower panel in
Fig. \ref{fig:tau} shows that when galaxies are selected within small
ranges of inclination, the attenuation-optical depth relation shows little scatter, with
the attenuation increasing as the inclination increases towards the edge-on
case.

Fig. \ref{fig:Dmagall} shows the median attenuation of model galaxies
at $z=3$, 6 and 8.2, as a function of both the intrinsic and attenuated UV absolute
magnitude. In both cases a trend can be seen for brighter galaxies
to be more attenuated, though this is dominated by scatter, especially
for attenuated UV magnitudes. The large scatter found suggests that,
in general, the attenuation is not related in a simple way with
the UV magnitude.

Fig. \ref{fig:Dmaglbg} is similar to Fig. \ref{fig:Dmagall} but for the
subsample of LBGs at $z=3$, 6 and 8.2, selected by applying (to the
attenuated colours) the magnitude and colour selection of
\citet{steidel95}, \citet{bou12} and \citet{lor11},
respectively. Fig. \ref{fig:Dmaglbg} shows that the range of absolute magnitude covered
by these galaxy selections is different. As was shown in Fig. \ref{fig:dropout}, the
LBGs selected by \citeauthor{steidel95} criteria are bright galaxies, with
$M_{\rm AB}(1500$\AA)$-5{\rm log}h\leq -20$. The selections
from both \citeauthor{bou12} and \citeauthor{lor11} make use of the
deep HST photometry, sampling fainter galaxies. In the next section we will explore these selections
further, comparing them with the model predictions.

Fig. \ref{fig:Dmaglbg} shows clearly that intrinsically brighter LBGs tend
to be more attenuated at all redshifts. However, when plotting against
the attenuated UV luminosity, the trend is weakened,
being dominated by the scatter. A direct estimation of the
  attenuation in high redshift galaxies spanning the range from 0 to
  5, similar to that predicted, has recently been made using results
  from {\it Herschel}
  \citep{burgarella11,buat12,lee12,reddy12}.

The right panel in Fig. \ref{fig:Dmaglbg} remains virtually unchanged if we only apply the cuts in
apparent magnitude from the observational surveys (see the
Appendix for the specific values). This suggests that the LBG colour selection does not lead to
an appreciable bias in average UV attenuation of the sample. In fact, the
right panel in Fig. \ref{fig:Dmaglbg} shows the complete overlap at
$z=3$ between the median attenuation of all galaxies and that for
LBGs, at least within
the magnitude range where the latter are selected.

\section{The predicted UV colours of LBGs}\label{sec:colours}
\begin{figure*}

\hspace{-0.5cm}
\begin{minipage}{5.8cm}
\includegraphics[width=6.4cm]{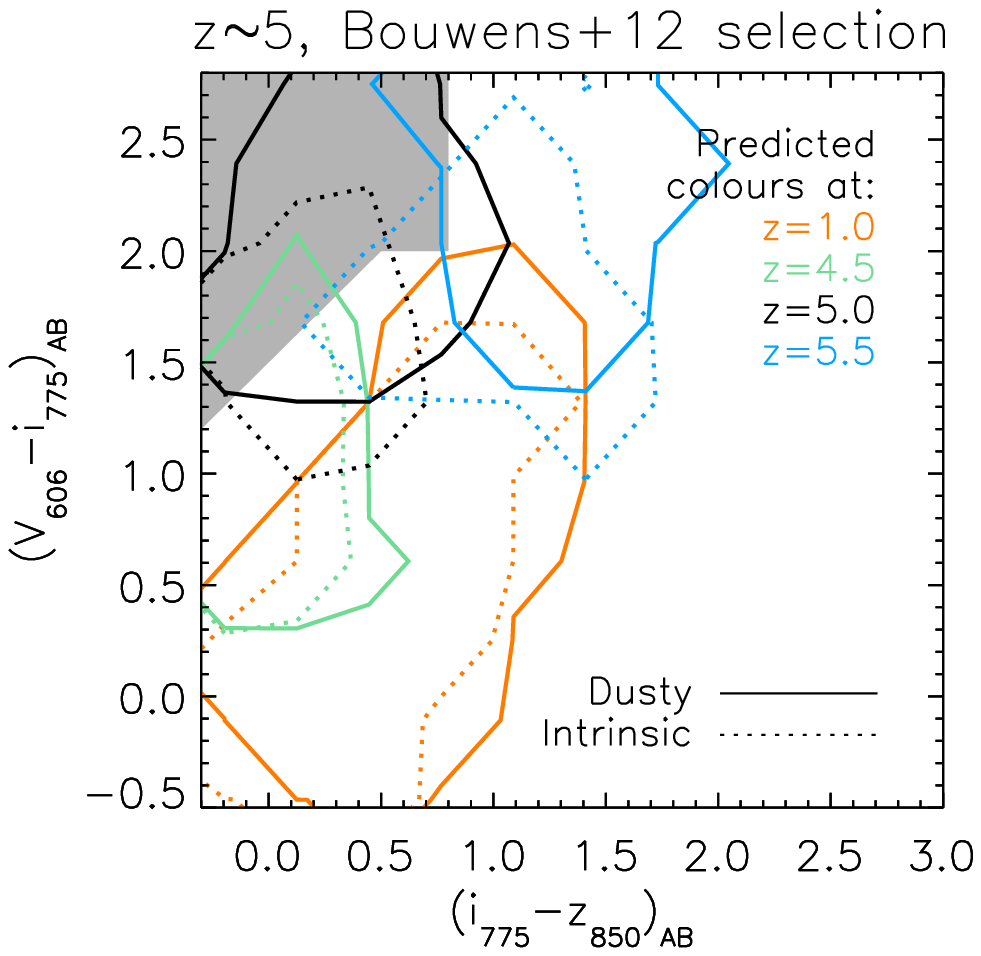}
\end{minipage}
\hspace{-0.5cm}
\begin{minipage}{5.8cm}
\includegraphics[width=6.4cm]{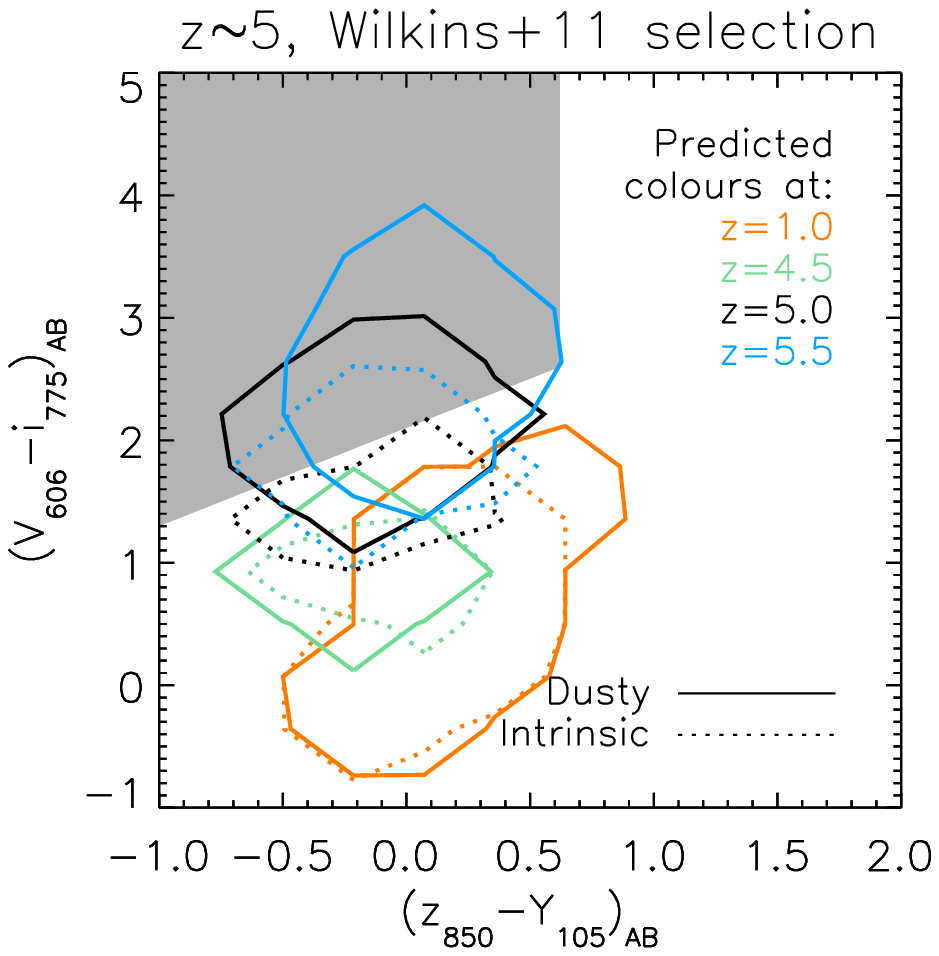}
\end{minipage}
\begin{minipage}{5.8cm}
\includegraphics[width=5.8cm]{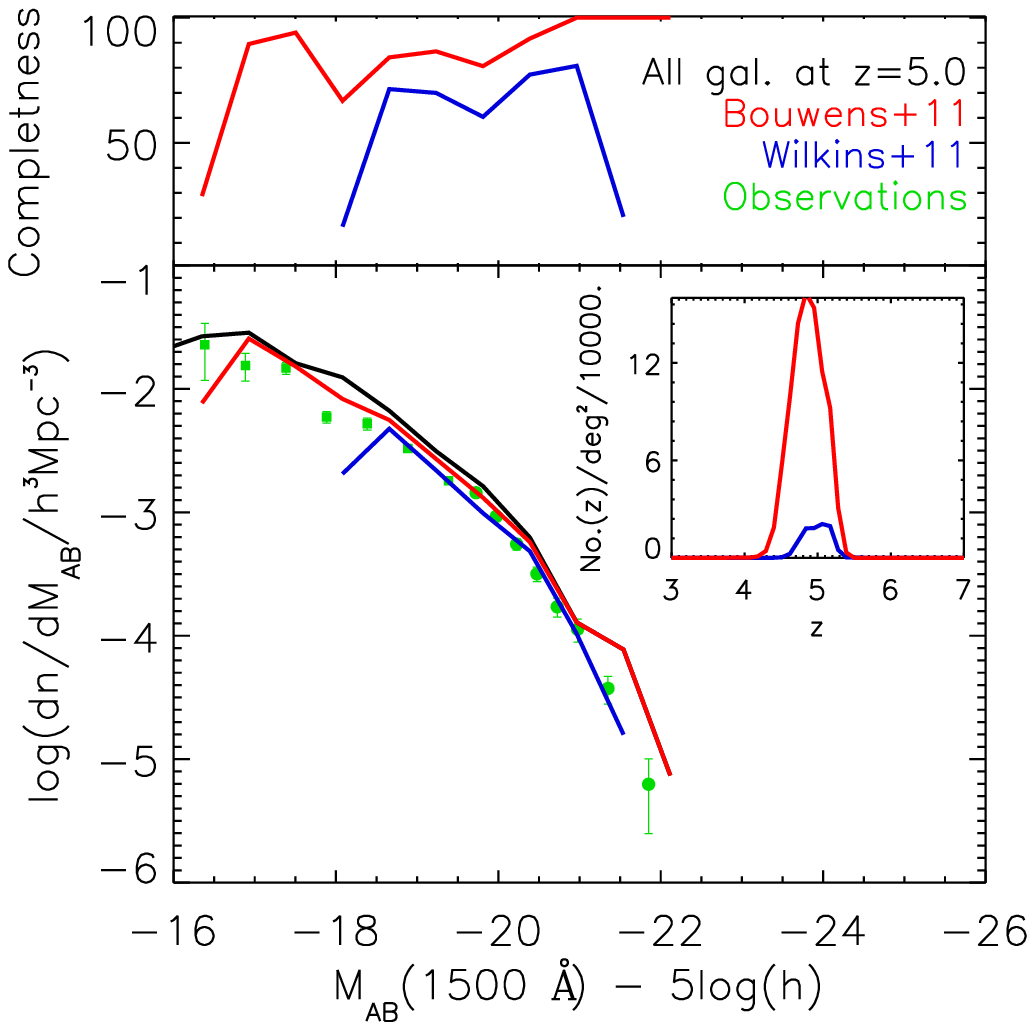}
\end{minipage}    

\hspace{-0.5cm}
\begin{minipage}{5.8cm}
\includegraphics[width=6.4cm]{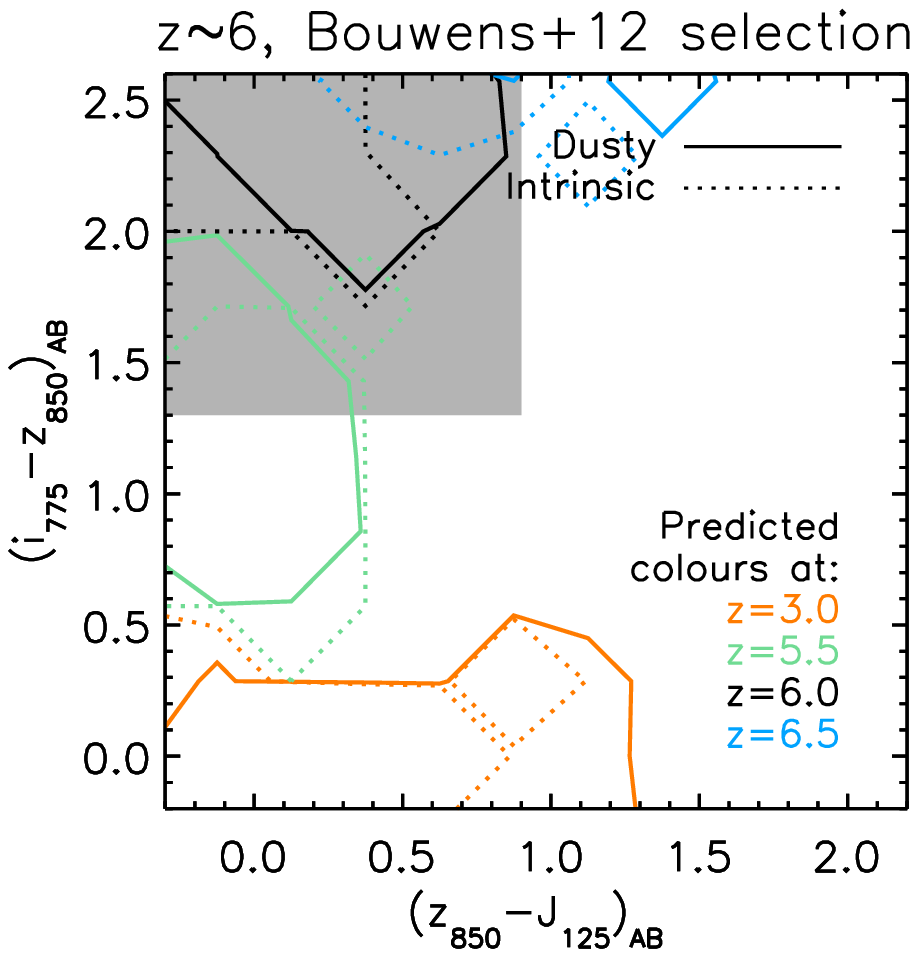}
\end{minipage}
\hspace{-0.5cm}
\begin{minipage}{5.8cm}
\includegraphics[width=6.4cm]{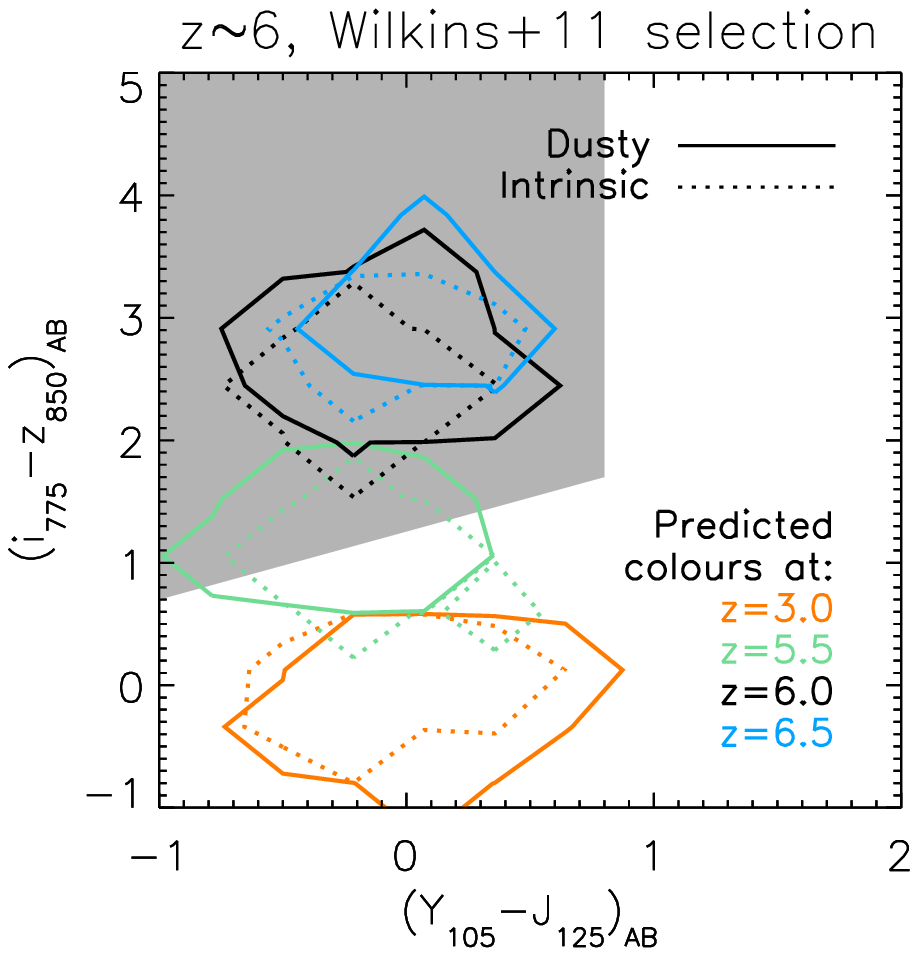}
\end{minipage}
\begin{minipage}{5.8cm}
\includegraphics[width=5.8cm]{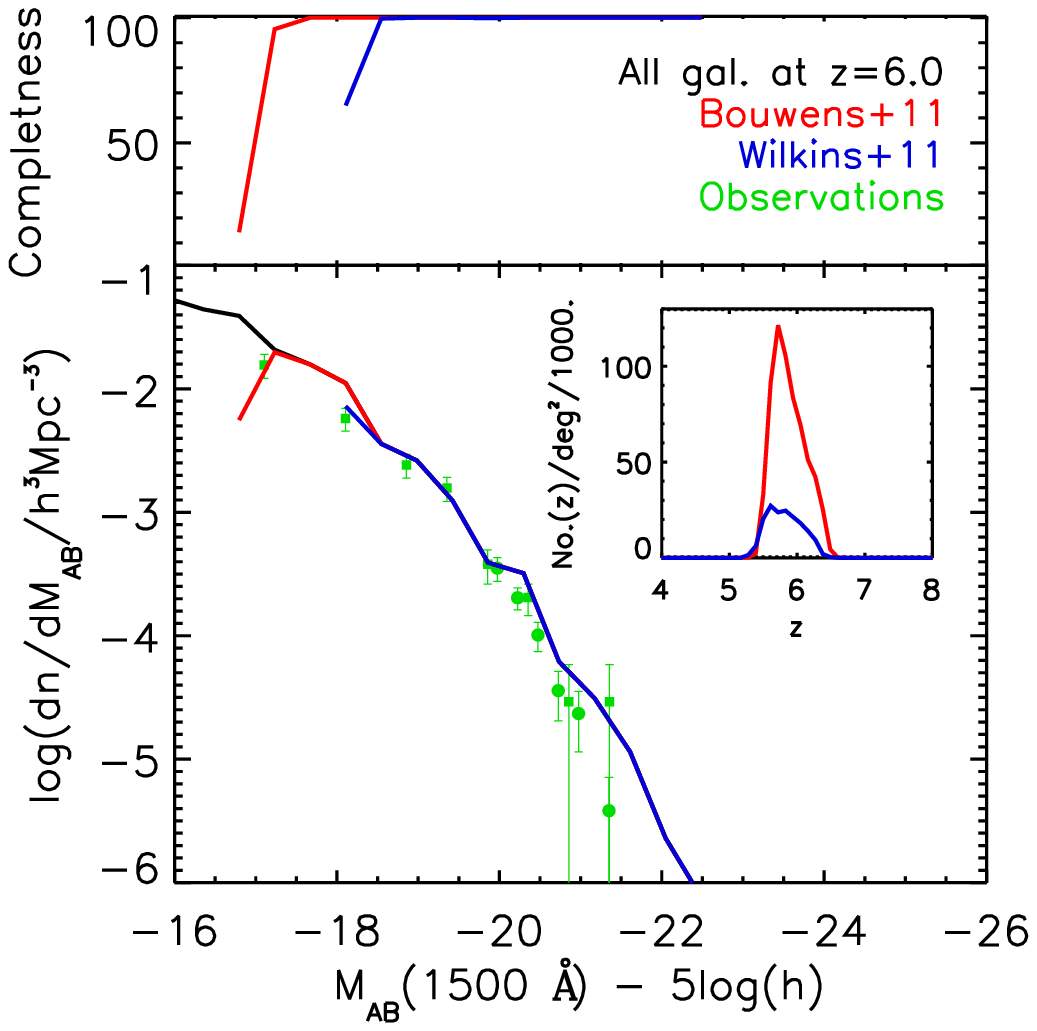}
\end{minipage}

\hspace{-0.5cm}
\begin{minipage}{5.8cm}
\includegraphics[width=6.4cm]{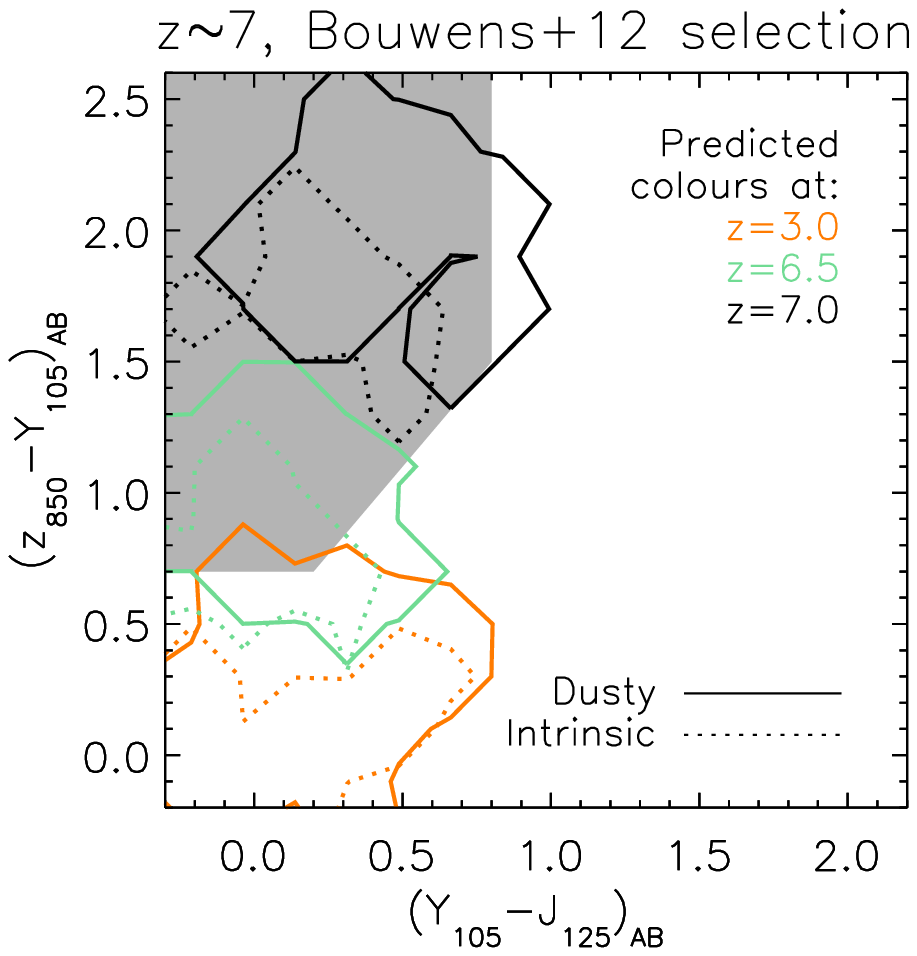}
\end{minipage}
\hspace{-0.5cm}
\begin{minipage}{5.8cm}
\includegraphics[width=6.4cm]{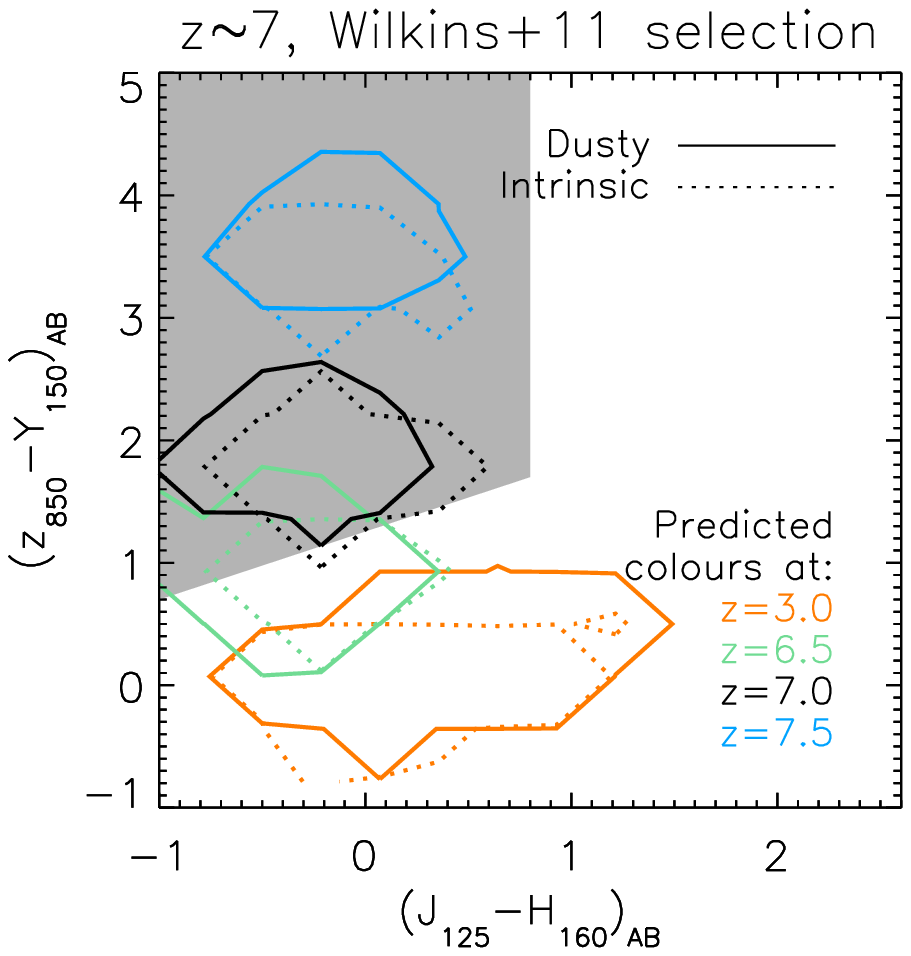}
\end{minipage}
\begin{minipage}{5.8cm}
\includegraphics[width=5.8cm]{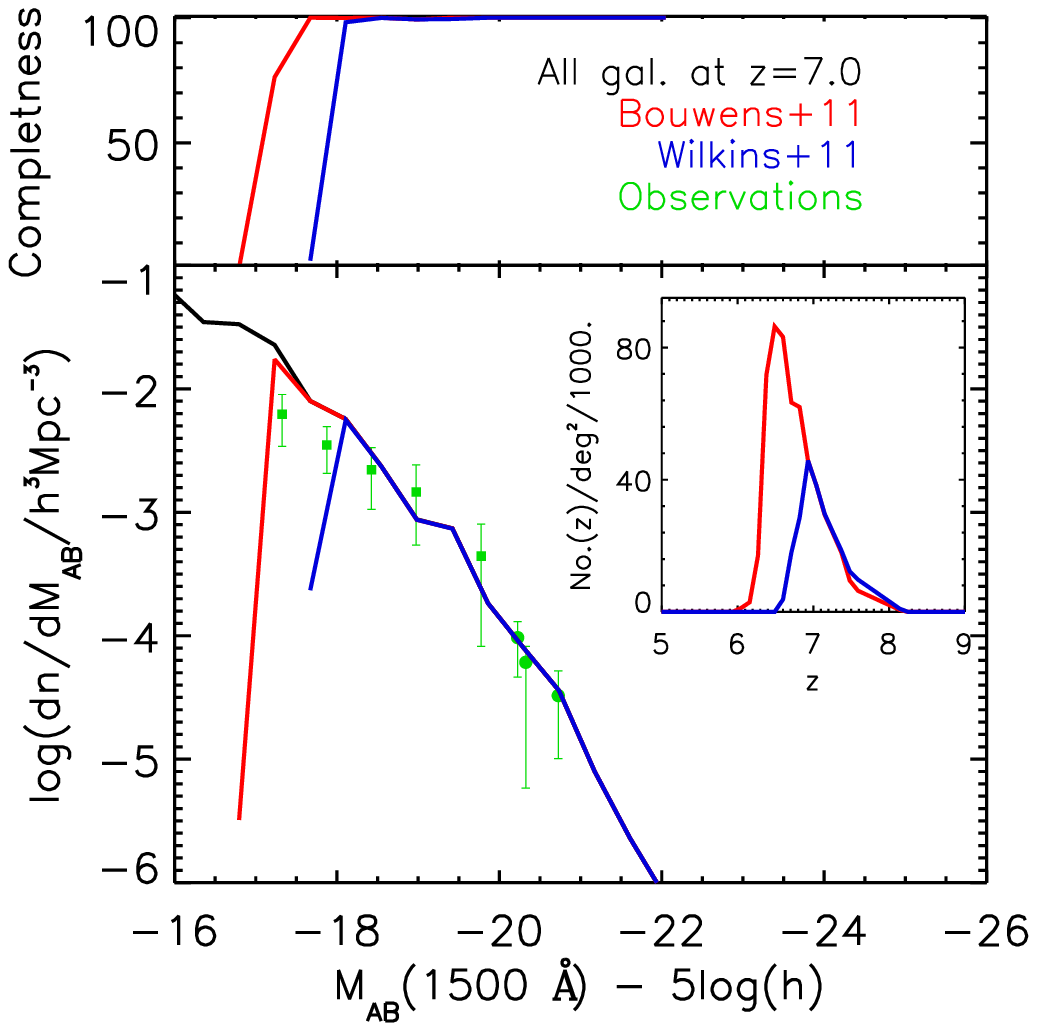}
\end{minipage}

\caption{The {\it left} and {\it central} columns show the predicted density
  contours for model galaxies at the redshifts indicated in the
  legend, in the colour-colour
  space defined by \citet{bou12} (left) and \citet{sw11} (centre)
  to select LBGs at $z=5$ (top row), $z=6$ (middle row) and $z=7$
  (bottom row) [grey areas in each plot]. The density contours enclose
  80 per cent of galaxies brighter than the limits set by the
  different observational studies (see the Appendix). The
  solid/dotted contours correspond to colours with/with out considering the
  dust attenuation. The
  {\it right} column shows as black lines the predicted completeness and luminosity function
  of all galaxies at these redshifts. The blue/red lines correspond to the predicted completeness
  and luminosity function for the subsample
  obtained using the selection criteria from
  \citet{sw11}/\citet{bou12}, the predicted redshift distributions of
  these subsamples are shown in the insets of the right column. For
  comparison, in the right column we also
  show in green the observed luminosity function from: 
   \citet{bou07} (squares, 1600\AA) and \citet{mclure09}
  (circles, 1500\AA) at $z=5$ (top);  \citet{bou07} (squares, 1350\AA) and \citet{mclure09}
  (circles, 1500\AA) at $z=6$ (middle);  \citet{mclure10} (squares, 1500\AA) and \citet{ouchi09}
  (circles, 1500\AA) at $z=7$ (bottom).
}
\label{fig:colcol}
\end{figure*}
\begin{figure*}

\hspace{-0.5cm}
\begin{minipage}{5.8cm}
\includegraphics[width=6.4cm]{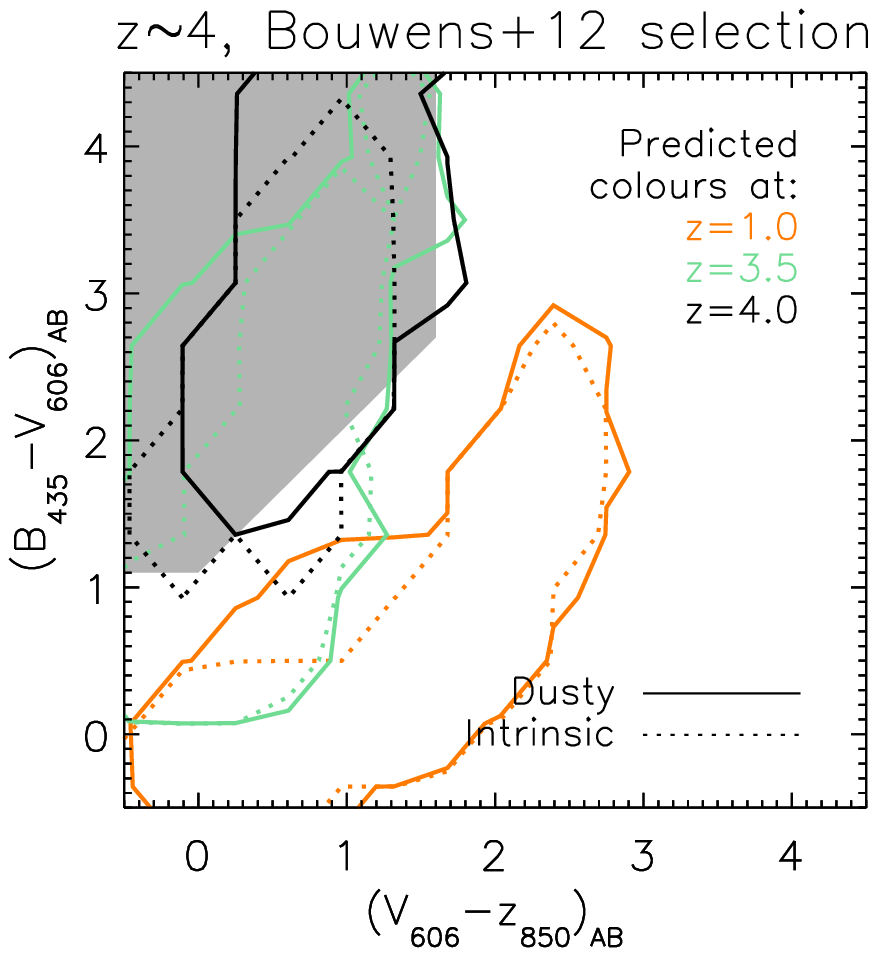}
\end{minipage}
\begin{minipage}{5.8cm}
\includegraphics[width=6.4cm]{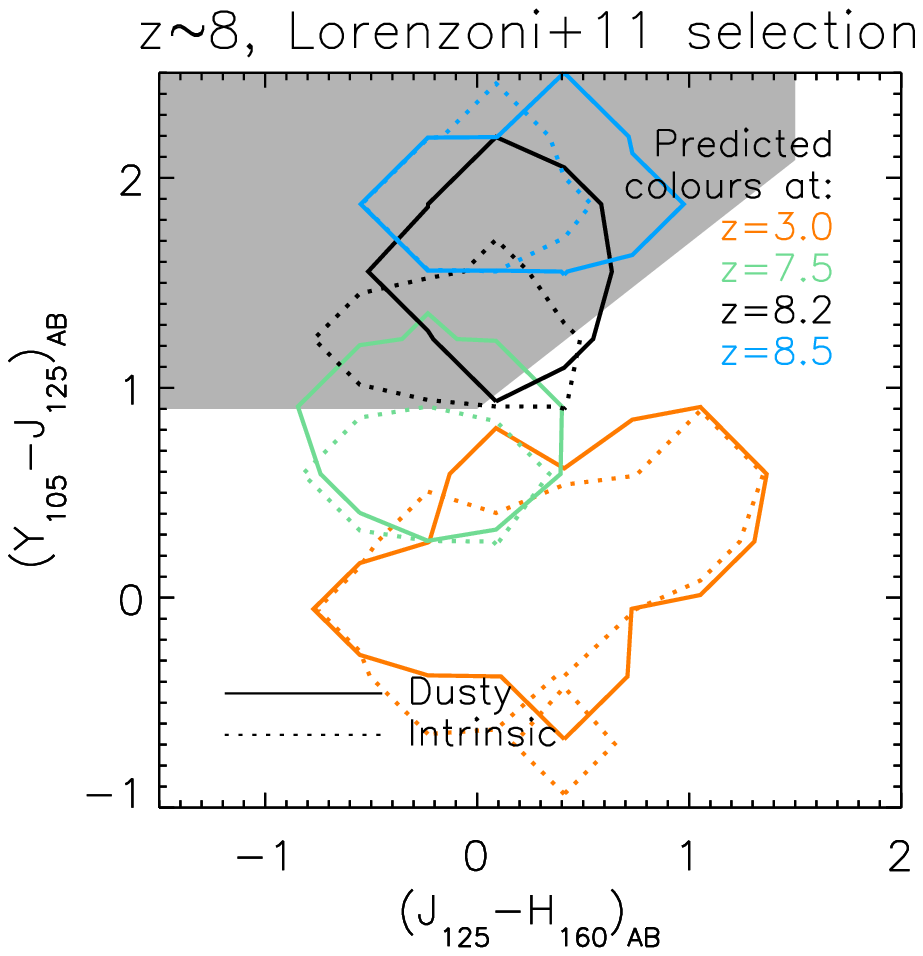}
\end{minipage}
\hspace{-0.2cm}
\begin{minipage}{5.8cm}
\includegraphics[width=6.4cm]{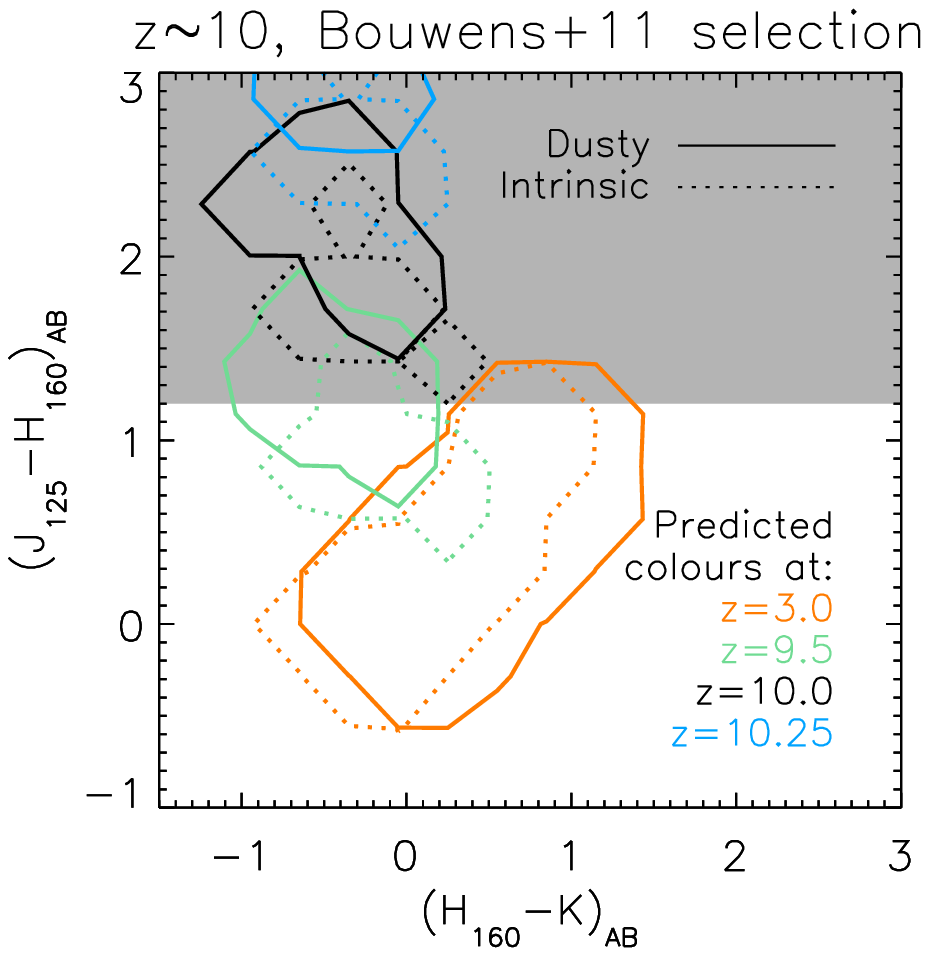}
\end{minipage}    

\begin{minipage}{0.5cm}
\hspace{0.5cm}
\end{minipage}
\begin{minipage}{5.6cm}
\includegraphics[width=5.6cm]{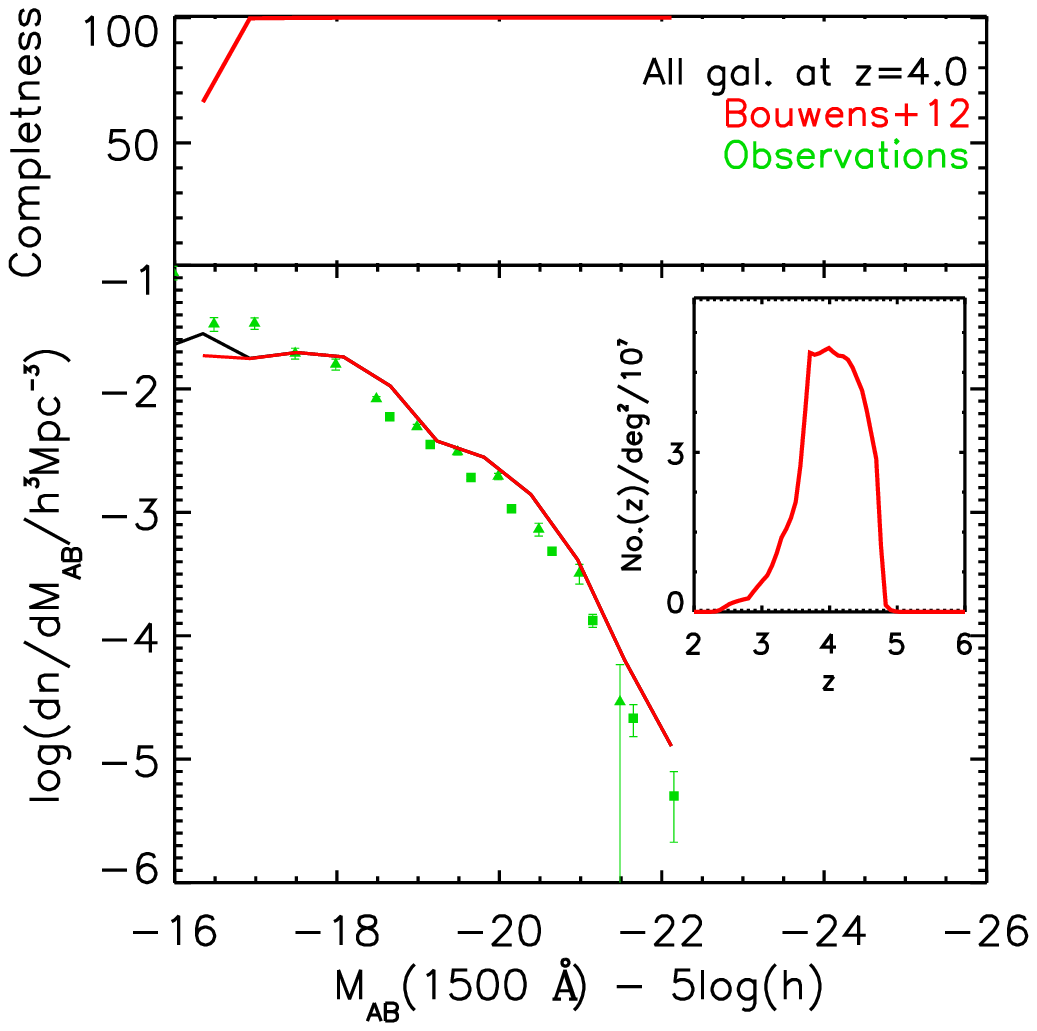}
\end{minipage}
\begin{minipage}{5.6cm}
\includegraphics[width=5.6cm]{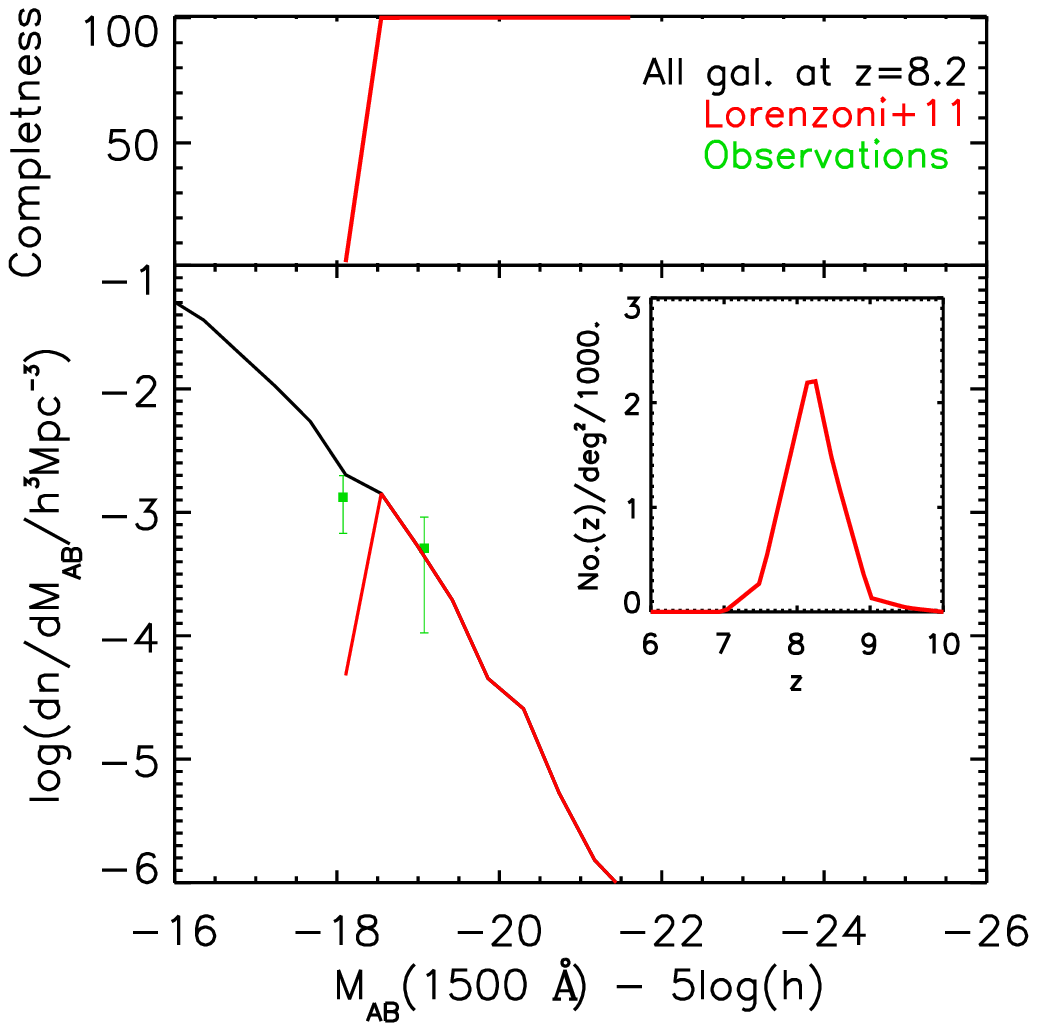}
\end{minipage}
\begin{minipage}{5.6cm}
\includegraphics[width=5.6cm]{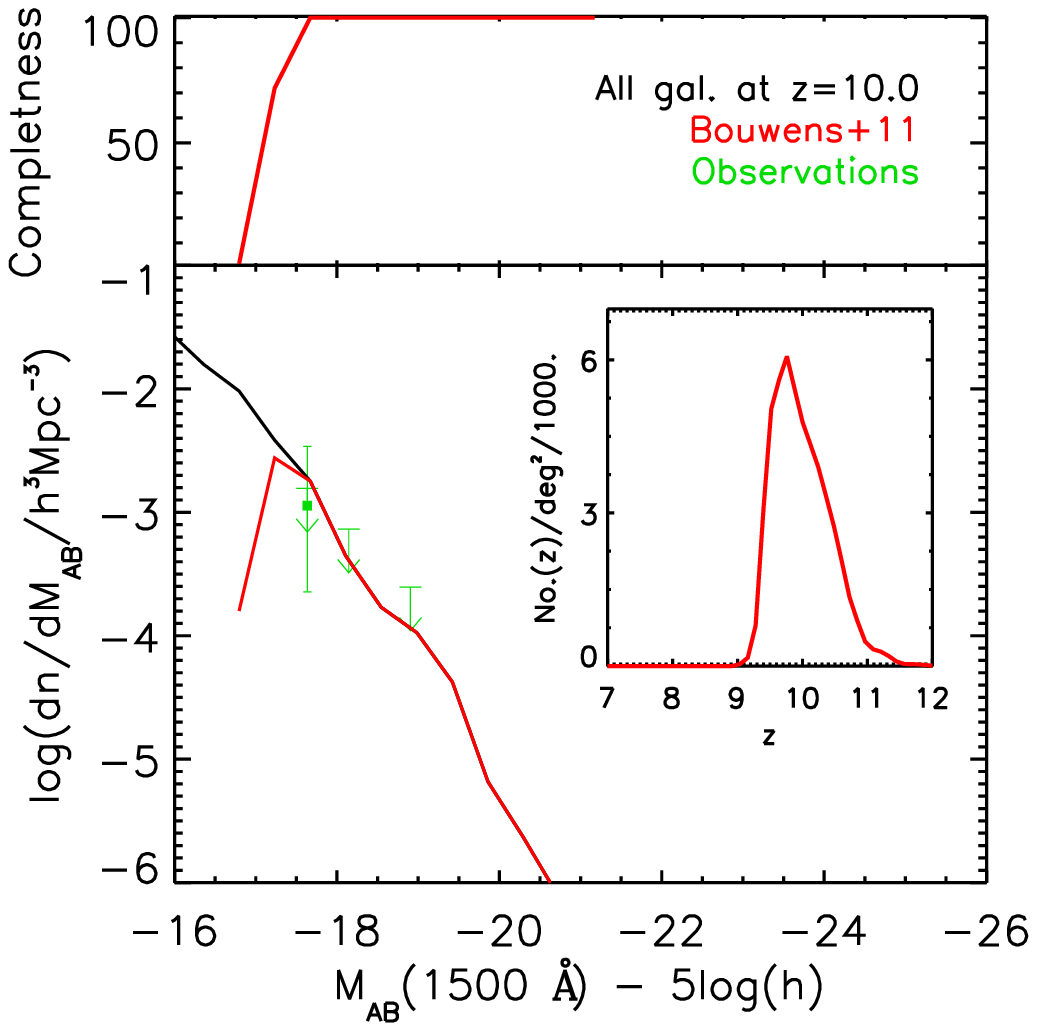}
\end{minipage}
\caption{{\it Top row:} The predicted density
  contours for model galaxies at the redshifts indicated in the
  legend, in the colour-colour
  space defined by the \citet{bou12} (left) \citet{lor11} (centre) and
  \citet{bou10} (right)
  to select LBGs at $z=4$, $z=8.2$ and $z=10$, respectively (grey areas in each plot). The density contours enclose
  80 per cent of galaxies brighter than the limits set by the
  different observational studies (see the Appendix). {\it Bottom
  row:} The black lines show the predicted completeness and luminosity function
  of all galaxies at  $z=4$ (left), $z=8.2$ (centre) and $z=10$ (right). The red lines correspond to the predicted completeness
  and luminosity function at the same redshifts for the subsample
  obtained using the selection criteria from \citet{bou12} (left), \citet{lor11} (centre) and
  \citet{bou10} (right). The predicted redshift distributions of
  these subsamples are shown in the insets (note that the y-axis units in
  the left panel are No.(z)/deg$^2/10^{7}$). For  comparison, we also
  show in green the observed luminosity function from:
  \citet{yoshida06} ($z\sim 4$, squares, 1500\AA, left), \citet{bou07}
  ($z\sim 4$, triangles, 1600\AA, left),  \citet{mclure10}
  ($z\sim 8.2$, 1500\AA, centre) and \citet{bou10} ($z\sim 10$, 1600\AA, right).
}
\label{fig:colcol2}
\end{figure*}

We start our study of the predicted rest-frame UV colours of LBGs by
comparing the locus occupied by model galaxies with the
colour region designed by observers to implement the
Lyman drop-out technique. Usually, these colour-colour regions are constructed by
following the colour evolution  of model starburst galaxies with
a range of negative UV
continuum slopes, predicted using a stellar
population synthesis (SPS) model. Figs. \ref{fig:dropout}, \ref{fig:colcol}
and \ref{fig:colcol2} show as grey areas some examples of such colour
selection regions\footnote{HST bands used here are F435W, F606W, F775W, F850LP, F105W, F125W, F160W (hereafter B$_{435}$,
V$_{606}$, i$_{775}$, z$_{850}$, Y$_{105}$, J$_{125}$, H$_{160}$,
respectively). The transmission curves of these filters can be seen in
Fig. \ref{fig:taus}.
}. In most cases, the observational data points are distributed
covering most of the drop-out region. In
these figures, we compare the observationally designed drop-out regions with the density contours for model galaxies brighter than the
limiting magnitudes in the corresponding observational studies (see
the Appendix for a list of the selection criteria used in different studies). As we can see in Figs. \ref{fig:dropout},
\ref{fig:colcol} and \ref{fig:colcol2}, the predicted UV colours of bright 
galaxies at $2.5\le z\le 10$ are consistent with the colour selection
regions designed observationally. 

In each of the colour-colour panels in  Figs. \ref{fig:dropout},
\ref{fig:colcol} and \ref{fig:colcol2}, the location of model galaxies at a range of redshifts, beyond the
target one, are shown to assesses whether or not there are
interlopers. In these figures we also present the predicted redshift distribution of the galaxies selected
following the magnitude and colour cuts from the
observational studies (see the Appendix). These predicted redshift distributions agree
with the target redshifts of each selection.

As can be seen in Figs. \ref{fig:dropout}, \ref{fig:colcol} and
\ref{fig:colcol2}, the predicted luminosity functions of  those galaxies selected following the application of
the magnitude and
colour cuts described in the different observational studies are in very good
agreement with observations up to $z\sim 10$. This is remarkable, since the
\citet{baugh05} model was designed considering only LBGs at
$z=3$. \citet{lacey11} presented a detailed study of the physical
processes affecting the predicted evolution of the UV luminosity function
and a comparison with observational results.

Fig. \ref{fig:colcol} shows that, with the HST filter set, it is particularly
difficult to select a complete sample of galaxies at $z=5$ whilst at
the same time attaining a minimum fraction of low
redshift interlopers. As previously reported by \citet{bou12}, the
expected locus of galaxies at $z=5$ in the colour-colour plot designed
for the drop-out technique partially overlaps that for galaxies
at lower redshifts. Thus, in order to minimise the fraction of low
redshift interlopers, the completeness of the high-z selection is
reduced. At $z=5$ we predict that the drop-out selection proposed by
\citet{sw11} recovers $\sim 70$ per cent of all galaxies with
$-21.5\leq$M$_{AB}$(1500\AA)$\leq -18.5$, and that from
\citet{bou12} $\sim 85$ per cent. Figs. \ref{fig:dropout},
\ref{fig:colcol} and  \ref{fig:colcol2} show that
the predicted completeness of the drop-out selection recovers more than
$90$ per cent of the galaxies brighter than the absolute
magnitude implied by the apparent magnitude observational limits  at $z=3$, 4, 6, 7, 8 and 10. Similar completeness
are also found for the predicted galaxies at $z=2.5$.

Next we explore how the predicted UV
colours change when different model parameters are varied. In
particular, we explore the effect of changing the stellar population synthesis (SPS) model and the
attenuation by both the IGM and dust.

\subsection{The effect of changing the SPS model}\label{col:IGM}
The \citet{baugh05} model uses by default the simple stellar populations
generated by \citet{bressan98}, using the Padova 1994
stellar evolutionary tracks\footnote{http://stev.oapd.inaf.it/cgi-bin/cmd, http://pleiadi.pd.astro.it/.} and the model stellar atmospheres from
\citet{kurucz93}\footnote{These are close to the isochrones and stellar atmospheres used by \citet{bc03}.}. We have compared the UV colours predicted using the
default SPS model with those from {\sc pegase.2} \citep{pegase}, \citet{mn} and
\citet{cw}. The SPSs models from both \citeauthor{mn} and
\citeauthor{cw} use stellar evolutionary tracks including a model for
the Thermally-Pulsating Asymptotic Giant Branch phase of stellar
evolution that can reproduce observations better than previous
attempts to model this stellar phase. \citeauthor{cw} also included a
post-AGB phase and an extended blue horizontal
branch for stellar evolution, matching UV observations of old star
clusters. 

The effect that stars in the Thermally-Pulsating Asymptotic Giant
  Branch phase have on the UV colours of a galaxy is negligible,
  since these stars are mainly bright in the near IR. Though bright in
  the UV, stars only reach a
  post-AGB phase at ages much older than those of the LBGs. Thus, little
  variation is expected in the UV colours predicted on using
  different SPS models. Indeed, we find that the UV colours of LBGs predicted using the different
SPS models are consistent with each other.

\subsection{The effect of the IGM on the UV colours}
\begin{figure}
{\epsfxsize=8.5truecm
\epsfbox[38 9 292 307]{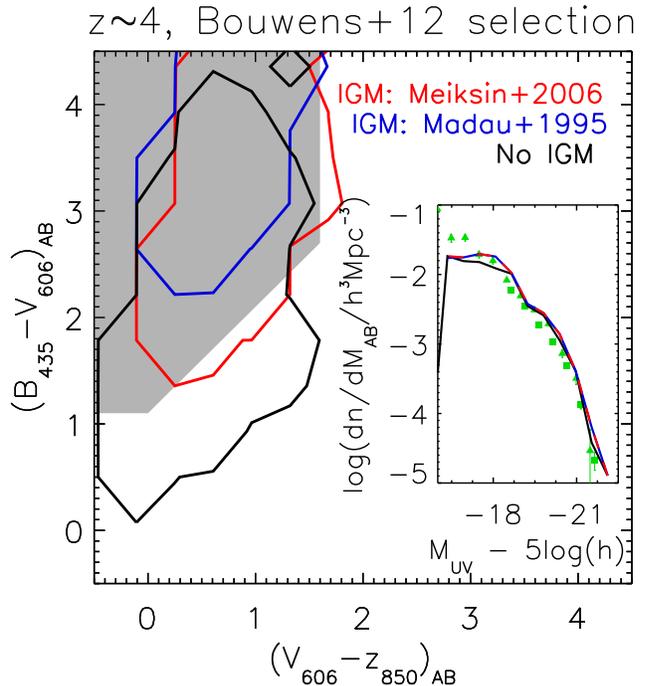}}
\caption{
The  (B$_{435}$-V$_{606}$)$_{AB}$ versus (V$_{606}$-z$_{850}$)$_{AB}$
  colour-colour plot proposed by \citet{bou12} for selecting
  galaxies at $z\sim 4$ (grey region) together with the predicted density
  contours for galaxies at $z=4$. The density contours enclose
  80 per cent of galaxies with V$_{606}<$30.1 and z$_{850}<$29.4. The inset shows the
  predicted luminosity function  for galaxies selected at $z=4$ with
  V$_{606}<$30.1, z$_{850}<$29.4
  and the colour cuts shown by the grey region in the colour-colour
  plot. In both cases, the black lines correspond to the predictions
  without considering the IGM
  attenuation and the blue/red lines to those predictions made using
  the \citet{ma}/\citet{me} prescriptions
  for the IGM attenuation. Note that the predicted luminosity
  functions using either IGM prescription are the same at this redshift. For comparison, we
  show in green the observed luminosity function from
   \citet{yoshida06} (squares, 1500\AA) and \citet{bou07} (triangles, 1600\AA).
}
\label{fig:igm}
\end{figure}

As described in
\S\ref{intro:IGM}, the IGM attenuates the UV light, mainly at
wavelengths $\lambda < 1216$\AA, and  has a larger effect at higher
redshifts.  Fig. \ref{fig:igm} shows as density contours the predicted
rest-frame UV colours for galaxies at $z=4$ with and without taking into account the
effect of the IGM attenuation. At $z\geq 3$,
the colours sampling the Lyman-break change by more than 0.5 magnitudes
when the IGM attenuation is included. In Fig. \ref{fig:igm} we can see
such change at $z=4$. It is clear the need to take into account the
effect of the IGM attenuation when studying the rest frame UV colours
of high redshift galaxies.

Fig. \ref{fig:igm} also shows the distribution of model galaxies when
either the \citet{ma} or the \citet{me} prescription is used to model the
IGM attenuation. In \S\ref{intro:IGM} we
pointed out that the prescription of \citeauthor{ma} shows a smaller
transmission than that of \citeauthor{me}. This implies that the predicted colours sampling the
Lyman-break are expected to be slightly redder for the model using the
\citeauthor{ma} prescription. Fig. \ref{fig:igm} illustrates this
difference for galaxies at $z=4$. Nevertheless, the difference in the
predicted rest-frame UV colours is found to be small. In fact, we do not find a
significant difference in the luminosity function of LBGs modelled
using either IGM attenuation prescription for galaxies at $2.5\leq
z\leq 10$ \citep[see the inset in
Fig. \ref{fig:igm} for the case of $z=4$ and LBGs selected following][]{bou12}.

In our model, reionisation is assumed to be instantaneous. Indeed, analyses of the line of sight to quasars at $z>6$ are
consistent with a uniformly ionised IGM at $z\sim
6$ \citep[e.g.][]{scho12}. At higher redshifts, numerical simulations of reionisation
suggest a gradual process, implying the existence of
ionised bubbles in a mainly neutral medium \citep{benson01,iliev06}.  The fact that the
predicted luminosity function including the effect of the IGM matches
the observations suggests that a minimal fraction of galaxies would
not be detected
due to the possible patchiness of the IGM at very high
redshifts.

\subsection{Dust treatment and the UV colours}\label{sec:dust}

In Figs. \ref{fig:colcol} and \ref{fig:colcol2} we compare the predicted distributions
of UV colours with and without considering the effect of dust. As can be seen there, in general, the presence of dust increases the spread of
UV colours. At $z=5$, the inclusion of dust clearly makes galaxies
redder in the colour sampling the Lyman-break for both the \citet{bou12}
and \citet{sw11} colour selections. At this redshift most
galaxies would be missed by these two drop-out techniques if intrinsic
colours were going to be used. As mentioned before, using the HST
  bands, the
selection of LBGs at $z=5$ needs to be done using stringent colour cuts,
in order to minimise low redshift interlopers. The effect of
dust on the UV colours at other redshifts, $z\neq 5$, is not so strong and
approximately the same completeness is recovered using the predicted
galaxy colours with and without considering the effect of dust.

Next we explore the effects that changing some of the model
parameters controlling the dust treatment have over the predicted
rest-frame UV colours.

\subsubsection{Changing the extinction law}\label{sec:ext}
\begin{figure}
{\epsfxsize=8.5truecm
\epsfbox[66 368 351 660]{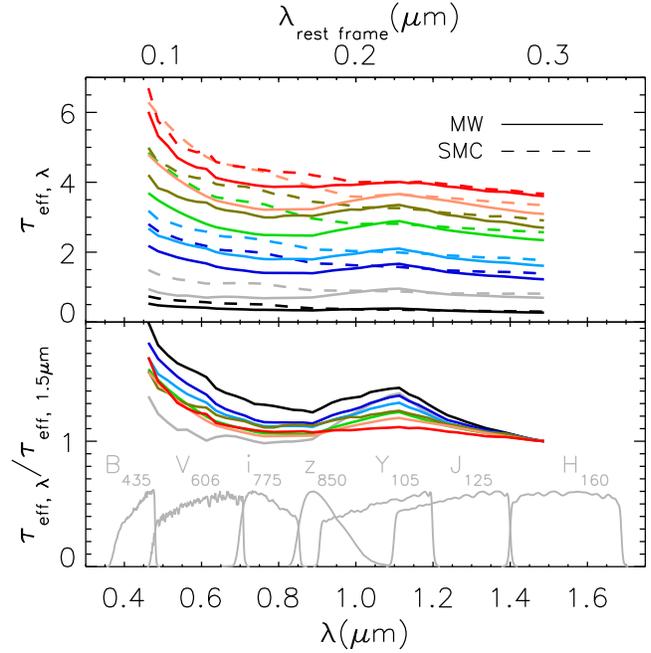}}
\caption
{{\it Top panel:} The optical depth as a function of wavelength for 8 galaxies at
  $z=4$ with different UV attenuations, obtained using 42 narrow
  top-hat filters starting with either the MW extinction curve (solid
  lines) or the SMC curve (dashed lines). {\it Bottom panel:} The
  optical depth normalised at $1.5\mu$m  for those galaxies in the top
  panel (each galaxy is presented as a line of the same colour in both
  panels). For reference, also shown
  are the response
  functions of the HST filters used to select LBGs at different
  redshifts (lower set of grey curves).}
\label{fig:taus}
\end{figure}
\begin{figure}
{\epsfxsize=8.5truecm
\epsfbox[28 5 292 281]{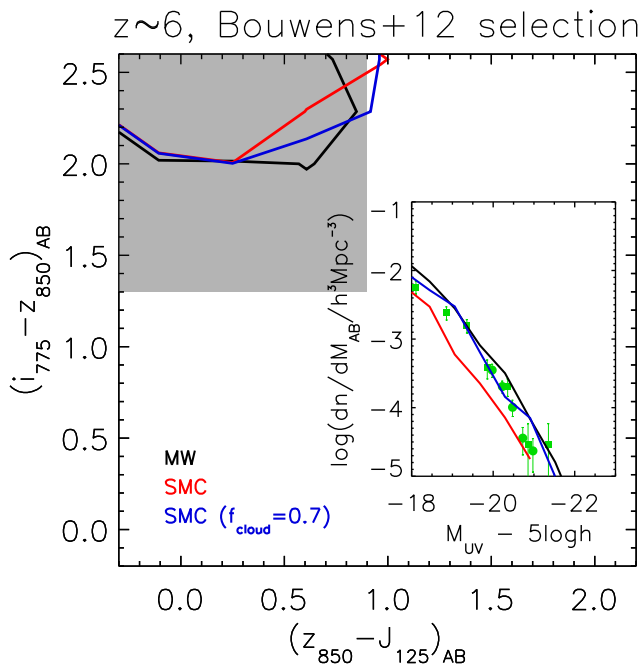}}
\caption
{The  (i$_{775}$-z$_{850}$)$_{AB}$ versus (z$_{850}$-J$_{125}$)$_{AB}$
  colour-colour plot proposed by \citet{bou12} to select
  galaxies at $z\sim 6$ (grey region) with the density
  contours for model galaxies at $z=6$. The density contours enclose 80 per
  cent of galaxies with z$_{850}<29.4$ and J$_{125}<29.9$. The inset shows the
  predicted luminosity function  for galaxies selected at $z=6$ with
  z$_{850}<29.4$ and J$_{125}<29.9$ and the colour cuts shown as a
  grey region in the colour-colour plot.  In both cases, the black lines
  corresponds to the default case of using the MW extinction law, the
  red to using the SMC extinction law and the blue lines corresponds
  to predictions made using the SMC extinction law and adopting
  $f_{\rm cloud}=0.7$ (where the default value is $f_{\rm cloud}=0.25$). For comparison, we
  show in green the observational data from  \citet{bou07} (squares, 1350\AA) and \citet{mclure09}
  (circles, 1500\AA).}
\label{fig:smc}
\end{figure}

The extinction curve of a galaxy
is determined by the composition and size distribution of the
galactic dust, which in
turn depends directly on the star formation history of a galaxy. At
high redshift it is particularly unclear which mechanism produced the
bulk of the observed dust
\citep{bianchi07,valiante09,mm10,dwek11,valiante11} and thus which
extinction curve best describes high redshift galaxies. Here, the dust attenuation calculation is based on the results of the radiative
transfer model of \citet{ferrara99}, which assumes a particular
extinction curve as an input.

Fig. \ref{fig:taus} shows for 8 galaxies, the effective optical depth as a function of
wavelength, obtained using
42 narrow, $50$\AA, top-hat filters starting with either the MW
extinction curve (the default one in the model) or the SMC extinction curve. These 8 galaxies
are at $z=4$, they are bright, $z_{850}>29.4$, they have inclinations
between 55 and 65 degrees and they have been
selected as they exhibit different attenuations in the $z_{850}$ band. 
We can see from this figure that the main features of both the MW and
SMC extinction curves, namely the bump at $2175$\AA\  for the MW curve and
the stronger UV extinction at smaller wavelengths in the SMC curve, are
still visible after the radiative transfer calculation, which takes into
account the distribution of dust in the galaxies and the viewing
angle. As we can see in
Fig. \ref{fig:taus}, the characteristics of the assumed extinction curve
affect the inferred dust attenuation. This result stresses the relevance of understanding better the
characteristics of dust distribution when interpreting observational
results. 

The lower panel in Fig. \ref{fig:taus} shows the effective optical
depth normalised to that at $1.5\mu$m, when starting with a MW
extinction curve. This panel shows clearly how
the features in the attenuation curves get smoother for higher
attenuations. The same is seen if we start with a SMC extinction curve
for dust in both the diffuse phase and in molecular clouds.

Fig. \ref{fig:smc} shows the distribution of model galaxies in the colour-colour plot proposed by \citet{bou12} to select
  galaxies at $z\sim 6$ (grey area) when different input extinction curves are used. The model galaxies populate the region
  designed to select galaxies at $z\sim 6$, independently of the input
  extinction curve. However, the median UV continuum slope is redder if we start with a SMC extinction law. We will return to this point in section \S\ref{sec:beta}. 

The inset in Fig. \ref{fig:smc} shows that the luminosity function of
LBGs at z=6 \citep[selected following][]{bou12}
predicted from the model using the SMC extinction law is about
half a magnitude fainter than that using the MW extinction law. The
luminosity function of all galaxies at $z=6$ starting the calculations
with either the SMC or the MW extinction curves still show this
difference and, thus, it cannot be due to the colour cuts applied in the
LBG selection. This difference is due to the fact that the SMC curve, as shown
in Fig. \ref{fig:att}, produces a larger attenuation in the rest-frame UV range
than is obtained from the MW curve. 

As discussed in \citet{lacey11}, the UV luminosity
function is very sensitive to the fraction of dust in molecular
clouds, $f_{\rm cloud}$, and the time that is assumed for stars to
migrate out of clouds, $t_{\rm esc}$. Increasing  $f_{\rm cloud}$ or
decreasing $t_{\rm esc}$ has the effect of reducing the dust
attenuation. By default, it is assumed that 25 per cent of the dust is
locked up in molecular clouds and that new stars will migrate out of these
after 1Myr. From MW observations, this escape time is found to be between 1 and 3 Myr \citep{tesc}. Thus, further reducing $t_{\rm esc}$ might
be unrealistic. However, increasing the amount of dust in clouds from 25 per cent could be
reasonable for some galaxies, in particular, for star forming
  galaxies at high redshift
 \citep[e.g.][]{daddi10}. The inset of Fig. \ref{fig:smc} shows
that assuming $f_{\rm cloud}=0.7$ increases the amplitude of the UV luminosity function
of LBGs modelled with the SMC extinction law, so it
gets closer to the prediction obtained using the MW extinction law and therefore to the
observational data. However, the predicted galaxies still populate a
similar region in the colour-colour plane shown in Fig. \ref{fig:smc}
and we find a UV continuum slope consistent with that obtained
starting with the default value of $f_{\rm cloud}=0.25$.

We have tested that changing either $f_{\rm cloud}$ or $t_{\rm esc}$
has a negligible impact on the predicted UV colours of model galaxies. However, the luminosity function depends on these parameters
and thus, the number of selected LBGs will be different for different
sets of parameters. This implies that UV colours cannot constrain
the dust parameters, $f_{\rm cloud}$ and $t_{\rm esc}$. A similar conclusion was reached by
\citet{lofaro09} in their study of model LBGs at $4\le z\le
6$.

\subsubsection{Changing the spatial distribution of dust}\label{sec:hz}

The way dust is distributed in a galaxy with respect to stars
affects the resulting attenuation and thus, it might affect the predicted
UV colours for a given galaxy. By default in {\sc galform} it is assumed
that the scale height of the dust distribution follows exactly that of
the stars, $h_{z,dust}/h_{z,stars}=1$. In order to explore how this
affects the predicted UV colours and luminosity functions of LBGs
at $3\le z\le 8.2$, we have made predictions for two other cases: when
dust is more concentrated than stars, $h_{z,dust}/h_{z,stars}=0.4$, and when it is more extended, $h_{z,dust}/h_{z,stars}=2.5$. 

The UV luminosity function is affected by the change in the relative
scale heights of dust and stars. When dust is more concentrated, it
affects the light of a smaller fraction of stars and thus the
galaxy presents a smaller attenuation compared to the default case of
having $h_{z,dust}/h_{z,stars}=1$. When the dust has a larger height
distribution than the stars, all of the starlight will be affected by
dust, as happens in the case of where $h_{z,dust}/h_{z,stars}=1$, and
thus, the two luminosity functions are very close. 

Despite the change
in the UV luminosity function, the model galaxies still populate
approximately the same region in the UV colour-colour planes.

\section{The UV continuum slope}\label{sec:beta}
\begin{figure}
{\epsfxsize=8.5truecm
\epsfbox[28 5 292 281]{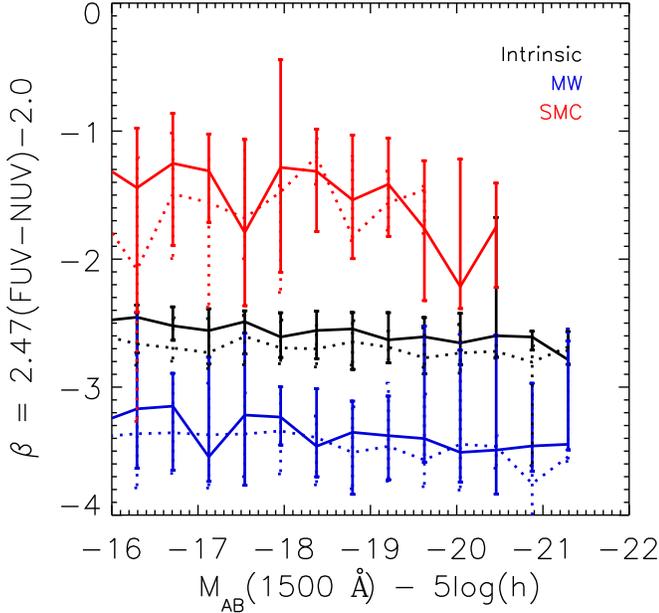}}
\caption
    {The UV continuum slope of model galaxies at $z=5$ (solid lines) and $z=7$ (dotted lines)
  measured with two top-hat filters, NUV ($\sim 1500$\AA )
  and FUV ($\sim 2200$\AA ), as a function of the  rest-frame UV absolute
  magnitude. The black line corresponds to
  the intrinsic UV continuum slope and the blue/red to that obtained
  using the MW/SMC extinction curve as the input for the dust attenuation calculation. 
}
\label{fig:b_ubd}
\end{figure}

The rest-frame UV continuum follows approximately a power law,
$f_{\lambda}\propto \lambda^{\beta}$, and thus, it can be characterised
by the slope $\beta$ \citep{meurer99}\footnote{Note that a $\beta=-2$
  corresponds to a galaxy with a flat spectrum in terms of $f_{\nu}$
  and, thus a zero UV colour in AB magnitudes.}. The UV continuum
slope is often considered to be directly related to the amount of dust
in a galaxy.  In \citet{swuvc} we studied the predicted intrinsic
(without attenuation) UV continuum slope, measured from a single colour. We
found that the distribution of intrinsic UV continuum slopes is affected by
the star formation and metal enrichment histories of galaxies and by
the choice of the IMF. These
dependencies limit the accuracy of the estimation of dust content in
galaxies from the observations of UV continuum slopes. We also found that the
intrinsic UV continuum slope is correlated with different galaxy properties,
including redshift, suggesting that the attenuation by
dust inferred from observations may be misestimated, if these
trends are not taken into account.

In most observational studies until now the UV continuum slope,
$\beta$, has been measured from a single UV
colour. Although this method is hindered by the photometric errors on
the two bands used, a fit to the spectral energy distribution of
galaxies at high redshifts is only now becoming possible \citep{finkelstein12,dunlop12}. Thus, in order to facilitate a comparison with
different observational studies, we calculate $\beta$ using a single UV colour. 

We have studied the impact on the UV continuum slope of modifying all the
model parameters discussed in \S\ref{sec:colours}. We find that the
main sensitivity is to the input extinction curve, which we discuss
further below. As expected from the previous sections, changing the SPS model does
not affect the predicted UV continuum slope. We have previously seen
that the IGM attenuation modifies the
UV colours, since this effect acts to enhance the strenght of Lyman-break. However, the
effect that the IGM attenuation has on the UV continuum slope
is negligible. We have studied how the UV continuum slope is modified
when the scale height of dust is different from that of
stars. As we saw in \S\ref{sec:hz} such a change mainly modifies the
UV luminosity function. Accordingly, if no magnitude cut is applied,
we find no variation in the UV continuum slope when we go from
$h_{z,dust}/h_{z,stars}=1$ to $h_{z,dust}/h_{z,stars}=0.4$ or to
$h_{z,dust}/h_{z,stars}=2.5$, in the studied redshift range. Similar
results are found when changing the parameters controlling the fraction of dust in clouds and the migration time out of clouds for young stars.

\subsection{The change of the UV continuum slope with luminosity}

Fig. \ref{fig:b_ubd} shows the median UV continuum
slope for galaxies at $z=5$ and
$z=7$. The UV continuum slope, $\beta$, has been obtained using:
\begin{equation}\label{eq:b}
\beta=2.47({\rm FUV-NUV})-2,
\end{equation} 
where (FUV-NUV) is the rest-frame UV colour calculated with two top hat filters covering the wavelength range: $1300\leq
\lambda _{FUV}$(\AA)$\leq 1700$ and $1800\leq
\lambda _{NUV}$(\AA)$\leq 2600$\footnote{Note that these two filters cover
  the same wavelength range as the GALEX filters.}.

As reported in \citet{swuvc}, our galaxy formation model predicts an
intrinsic UV continuum slope that is practically independent of the UV luminosity
of galaxies.  We find that the predicted UV continuum slope varies by
less than 0.2 magnitudes
for $-22\le M_{\rm UV}-5{\rm
  log}h\le-16$. We find an intrinsic UV continuum slope of about $-2.5$ for galaxies $5\le z \le
7$. Similar values have been found in hydrodynamical
simulations developed to study LBGs at $z>5$
\citep{jaime10,pratika12}. 

Fig. \ref{fig:b_ubd} also shows the predicted UV continuum slope once
the effect of dust is taken into account, starting from either the MW
or the SMC extinction curves. We find that the UV continuum slope is
very sensitive to the extinction curve used as input for the dust
attenuation calculations. For the chosen set of filters, we find that,
as expected, galaxies get redder when the dust attenuation is
calculated starting with the SMC extinction curve. However, as shown in Fig. \ref{fig:b_ubd} model
galaxies get {\it bluer} when the dust attenuation is
calculated starting with the MW extinction curve. 

\begin{figure*}
\begin{minipage}{5.8cm}
\includegraphics[width=5.8cm]{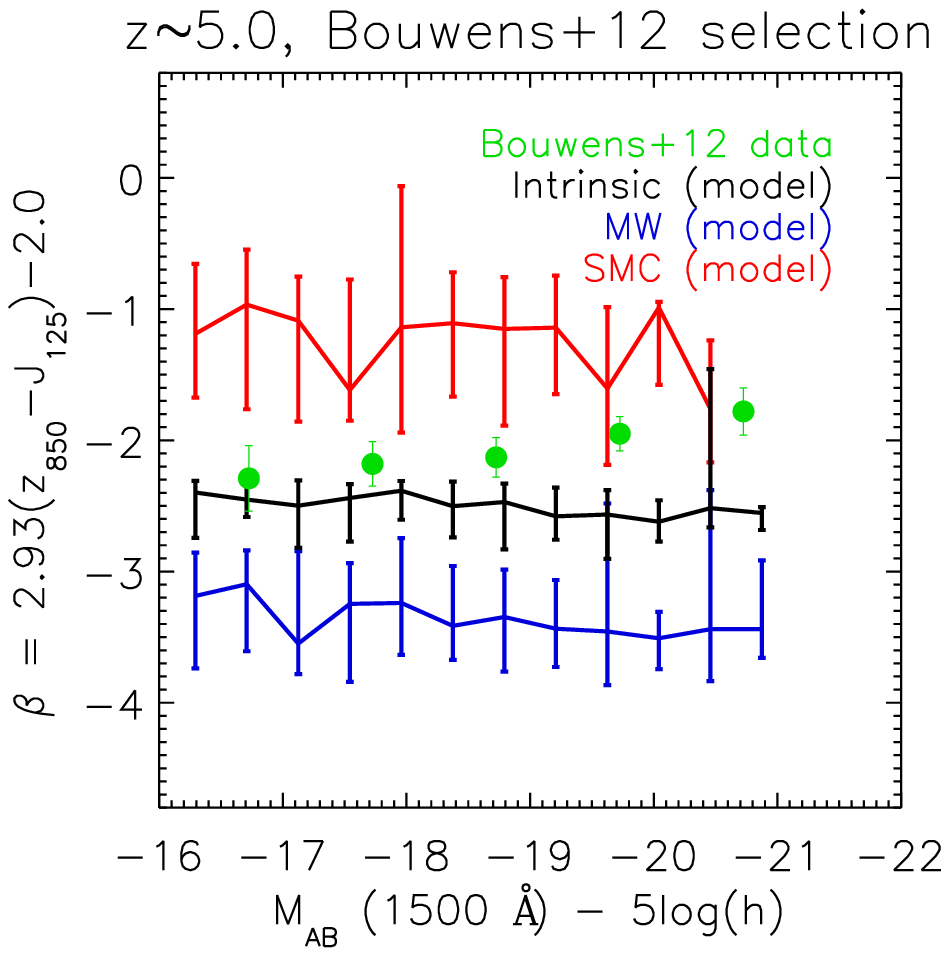}
\end{minipage}
\begin{minipage}{5.8cm}
\includegraphics[width=5.8cm]{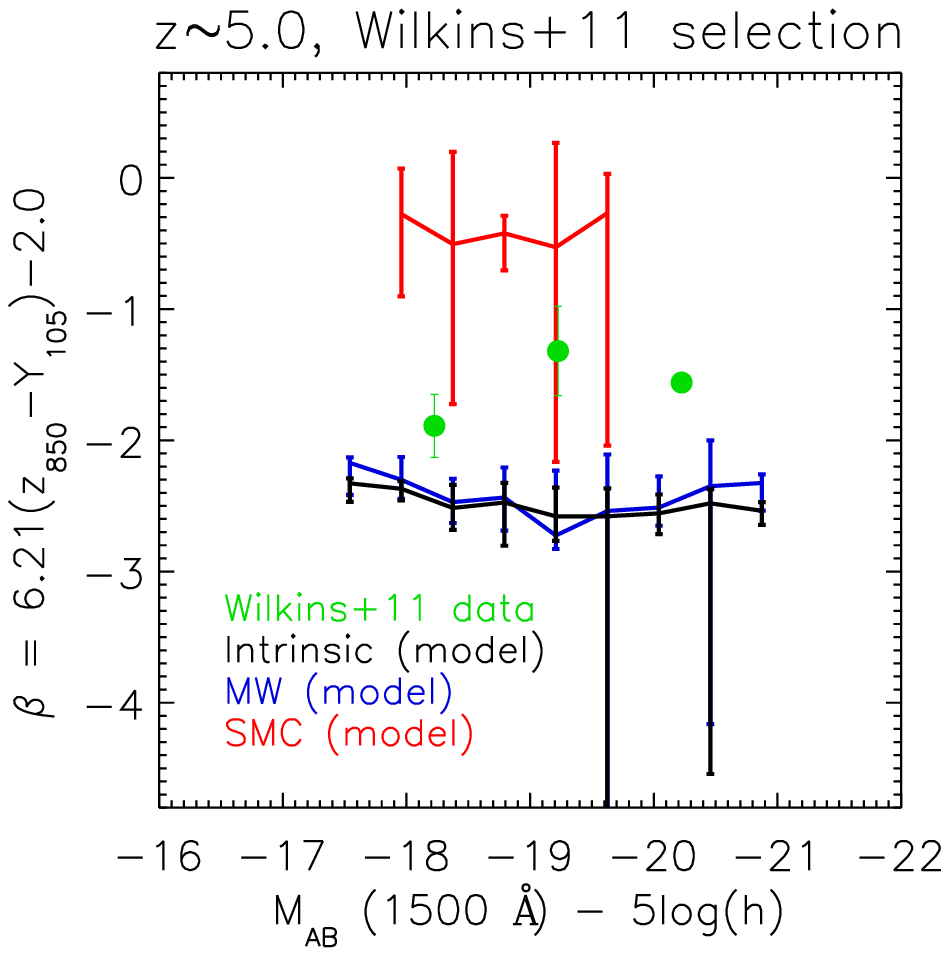}
\end{minipage}
\begin{minipage}{5.8cm}
 \hspace{5.8cm}
\end{minipage}

\begin{minipage}{5.8cm}
\includegraphics[width=5.8cm]{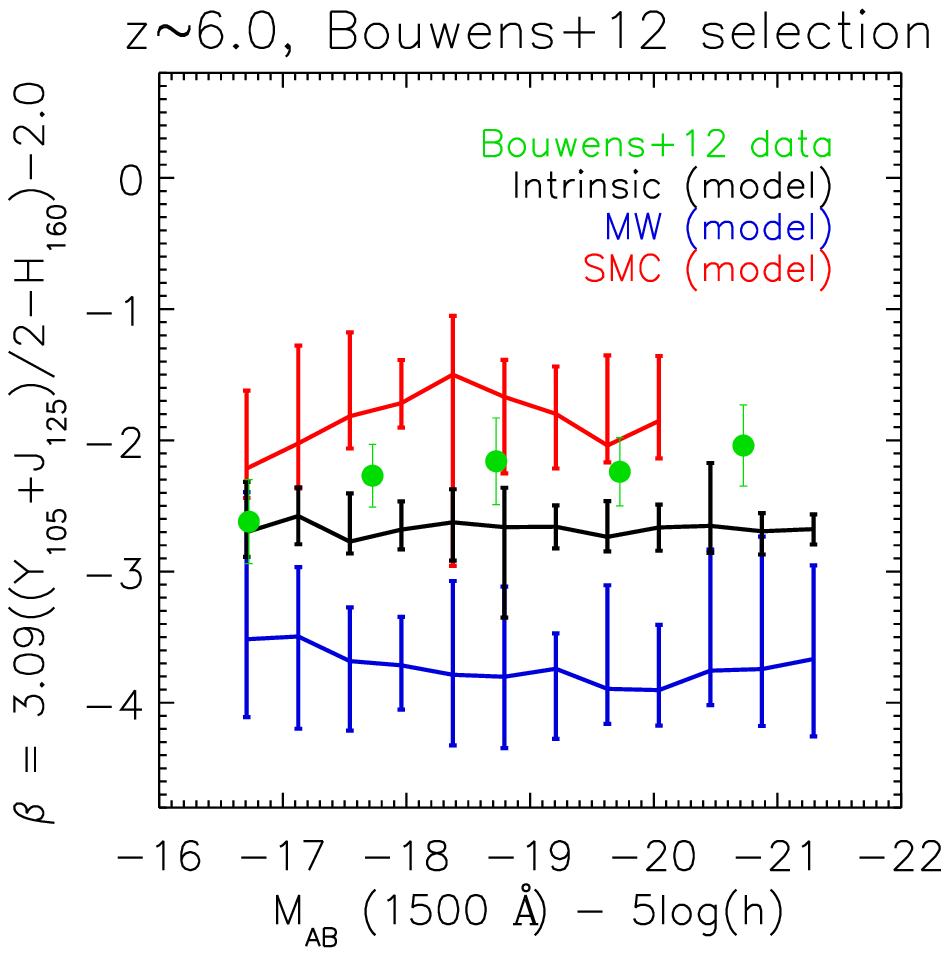}
\end{minipage}
\begin{minipage}{5.8cm}
\includegraphics[width=5.8cm]{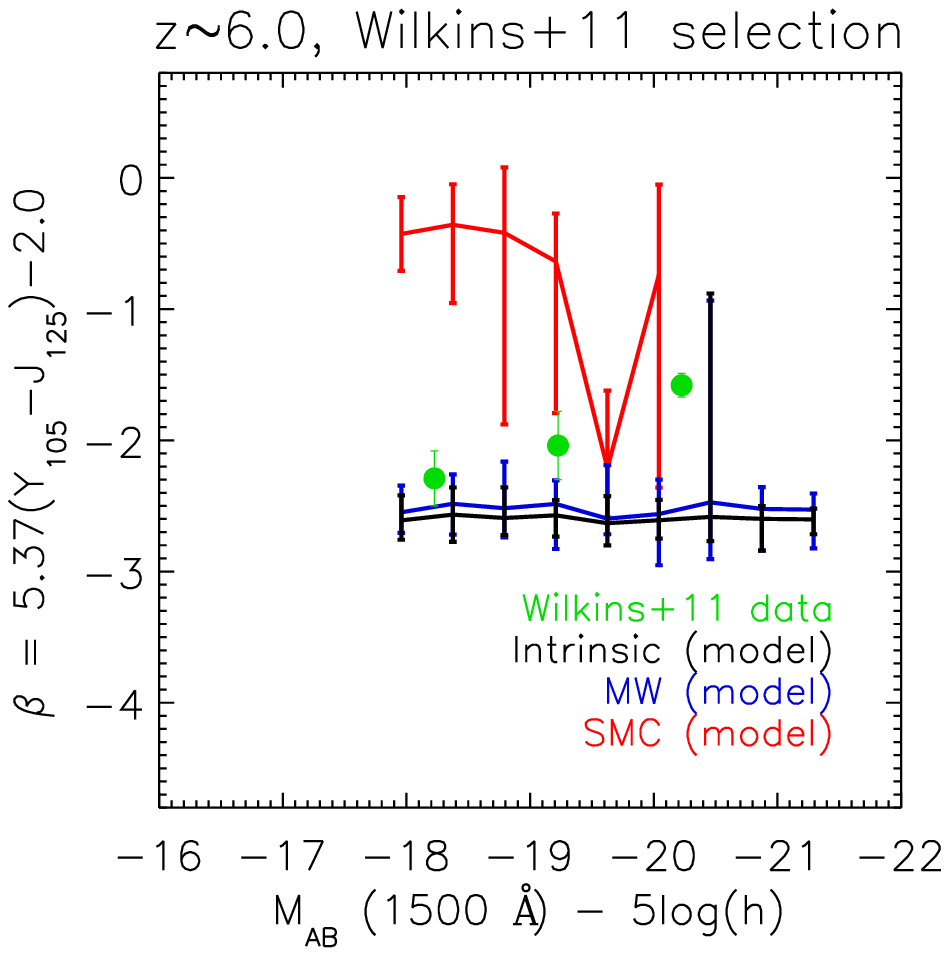}
\end{minipage}
\begin{minipage}{5.8cm}
\includegraphics[width=5.8cm]{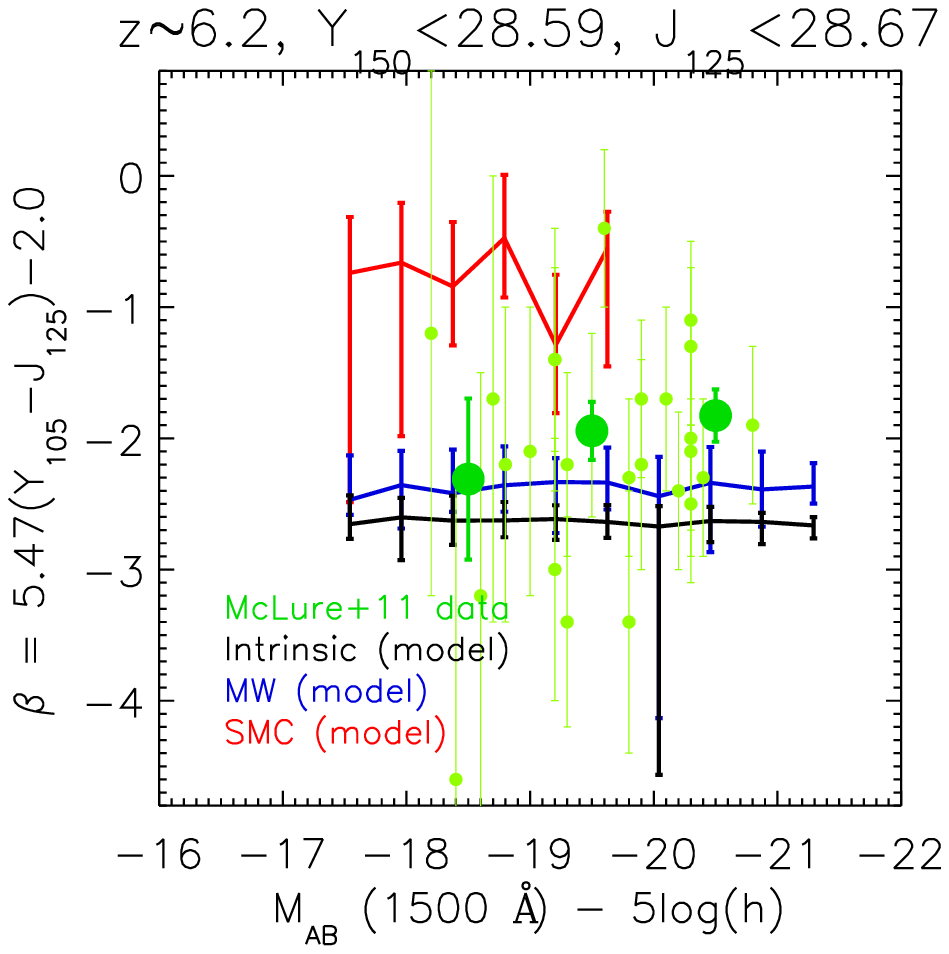}
\end{minipage}

\begin{minipage}{5.8cm}
\includegraphics[width=5.8cm]{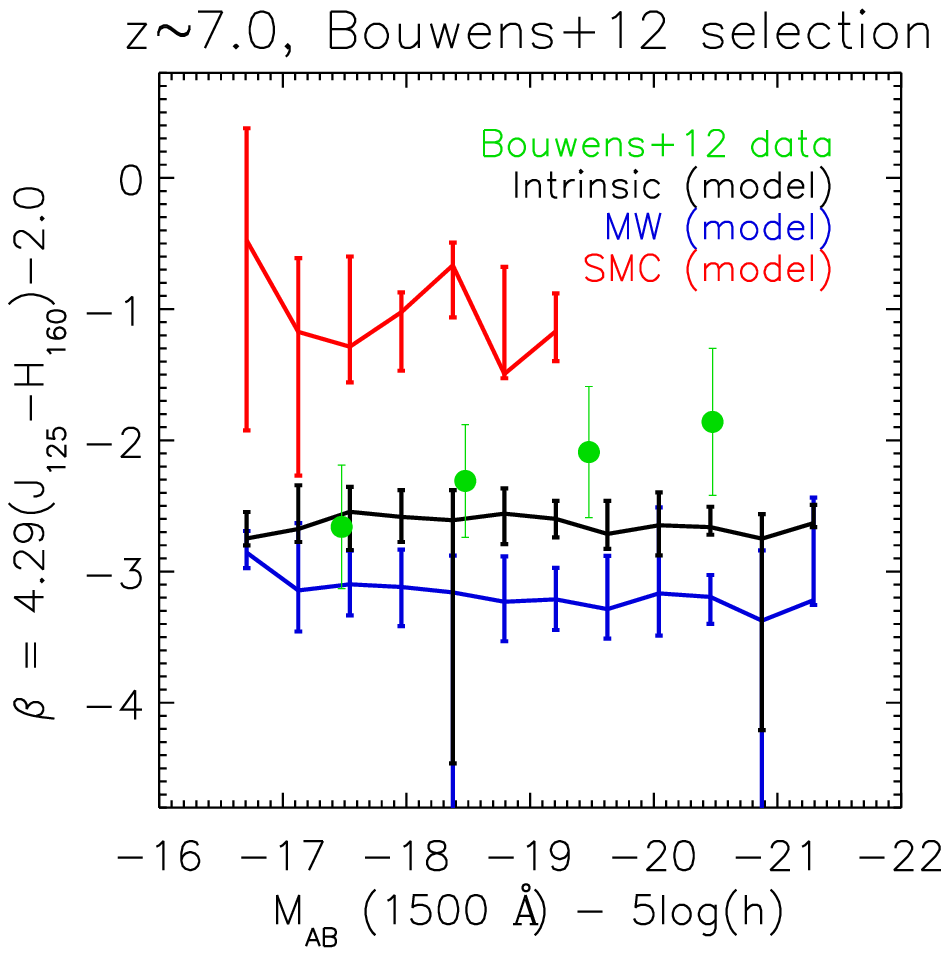}
\end{minipage}
\begin{minipage}{5.8cm}
\includegraphics[width=5.8cm]{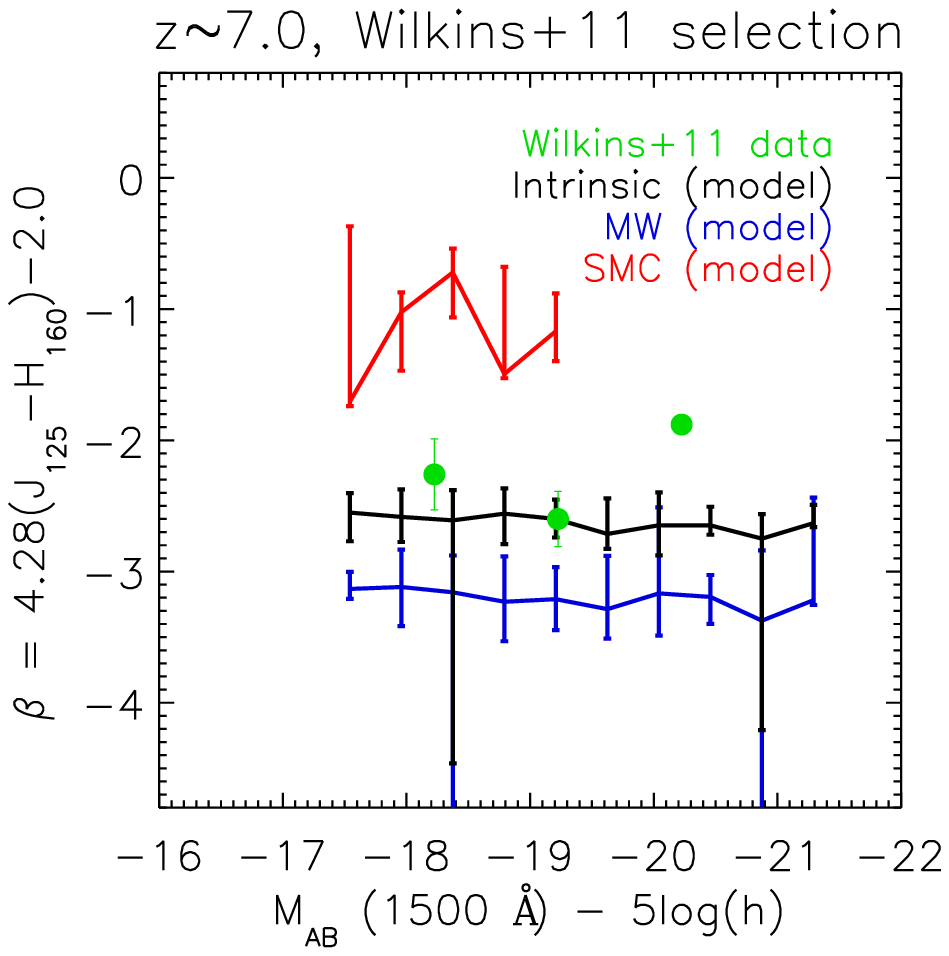}
\end{minipage}
\begin{minipage}{5.8cm}
\includegraphics[width=5.8cm]{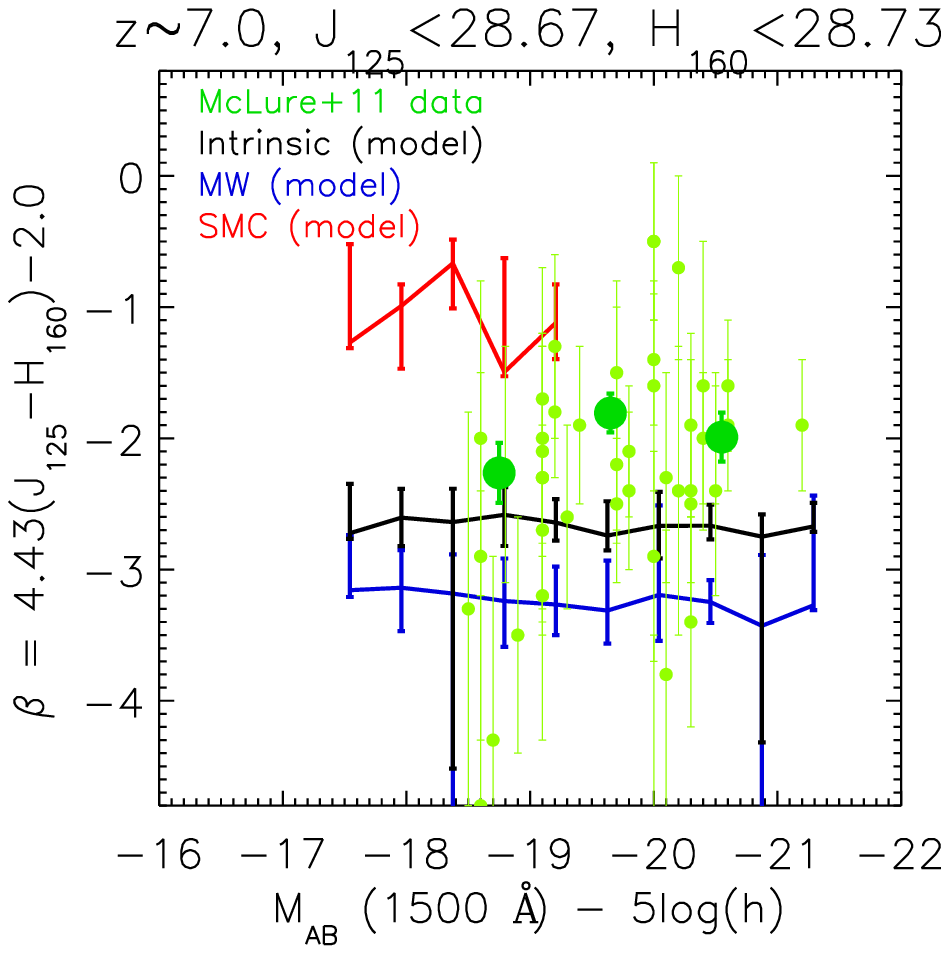}
\end{minipage}

\caption{The predicted median intrinsic UV continuum slope (black
  lines) and the attenuated slope, starting either with the MW (blue lines) or the
  SMC extinction curve (red lines), as a function of absolute
  attenuated UV
  magnitude, for galaxies at $z\sim 5$ (top), $z\sim 6$
  (middle row), $z\sim 7$ (bottom), selected following the magnitude and
  colour cuts proposed in different observational studies (summarised in
  the Appendix). The different columns correspond to the UV continuum
  slope calculated following the method adopted in the observational studies indicated on
  the top of each plot. The error bars show the 10 and 90 percentile range of the
  model predictions. The green circles present, for comparison, the
  observational data from \citet{bou12} (left column), \citet{sw11}
  (middle column) and \citet{mclure11} (right column, who selected LBGs
  based on photometric redshift and not UV colour cuts). In the latter case
  we also show the observations for individual galaxies in light
  green. The mean observational points are shown with their associated errors.}
\label{fig:beta}
\end{figure*}

This behaviour is directly related to the $2175$\AA\  bump in the MW extinction curve shown in Fig. \ref{fig:att}, that is
still present once the geometry of the dust is taken into account, as
seen for the 8 example galaxies in Fig. \ref{fig:taus}. Thus, this behaviour
will depend on the particular set of filters chosen to
measure the galaxy colour and it is expected to be reduced if a SED
fitting is used to calculate the UV continuum slope. Both \citet{jaime10} and \citet{pratika12} estimated the UV
continuum slope by SED fitting their model galaxies, finding very
good agreement with the observed values. These two studies estimated the
dust attenuation assuming the simple approximation of a dust
slab. \citet{pratika12} assumed that the dust present in galaxies at
$z>6$ is produced solely by Type-II SNe. The extinction curve of dust
produced solely by SNe presents a bump at $\sim 2500$\AA\ and a very
steep slope in the UV \citep{bianchi07}. These characteristics could
also imply that galaxies with dust solely produced by SNe will appear to having bluer UV
continuum slopes if they were estimated using a single colour.

\subsubsection{Comparison with observations}

In order to directly test our predictions
against observations, we calculate $\beta$ using the exact filters and
equations proposed in different studies that use a single UV colour to
estimate the slope of the UV continuum. Fig. \ref{fig:beta} shows the predicted intrinsic UV continuum slopes
calculated following different observational studies for galaxies at
$z\sim 5,$ 6 and 7, using the magnitude and colour cuts summarised in
the Appendix. 

At $z=6$, Fig. \ref{fig:beta} illustrates how,
when starting from the MW extinction curve, model galaxies can look redder,
the same or even {\it bluer} than their intrinsic colours, depending on the
choice of filters. This is directly related to the 2175\AA\ bump being
sampled or not by the two chosen filters. This extreme
sensitivity of $\beta$ to the choice of colour used in the estimation
reinforces the results from  \citet{finkelstein12}, who showed how
weakly the UV continuum slope is constrained with a single colour
compared with SED fitting using at least 5
bands. 

The right column of Fig. \ref{fig:beta} shows the UV continuum slopes
obtained for individual galaxies in the \citet{mclure11} sample, together
with the mean values per magnitude bin, as shown for the
samples from both \citet{bou12} and \citet{sw11} in the other
columns of Fig. \ref{fig:beta}. The scatter when considering individual galaxies is large
enough to cover the UV continuum slopes predicted starting with both
the MW and SMC extinction laws. However, \citet{mclure11} presented a
detailed analysis of the problems that affect individual
measurements and concluded that averaged UV
continuum slopes are more reliable. 

Fig. \ref{fig:beta} shows that, for some filter combinations, the predicted $\beta$ including the effect of dust agree
with the observations. However, in most cases, the observed mean values are
closer to the predicted intrinsic values than to those including the
effect of dust. 

By doing the exercise of smoothing out the 2175\AA\ bump in the 
attenuation curves calculated with the \citet{ferrara99} model (which
are still a function of inclination, optical depth, etc.), we obtain
UV continuum slopes about 0.2 magnitudes redder than the intrinsic ones
and which are therefore closer to, and in some cases consistent with,
the observed values. We have also used the attenuation at 1500\AA\
obtained in the above calculation to normalise the application of the
Calzetti law to the model galaxies to derive another illustrative estimate of the
attenuation. We find that the attenuation curves from this procedure
are greyer than the Calzetti law i.e. there is less wavelength
dependence. By using the Calzetti law normalised in this way, we obtain attenuated
UV continuum slopes which are consistent with those inferred
observationally in at least one magnitude bin at z=5, 6 and 7. Therefore, assuming the Calzetti
law, observations are consistent with the predicted 1500\AA\ attenuations.

The predicted values of the UV continuum slope
are extremely sensitive to the calculation of the dust attenuation. The observational data are well within the range of predictions when
considering either the MW or the SMC extinction curves as input,
again emphasising the uncertainty in the treatment of attenuation, and the
implications that has when interpreting observational data. It is
unclear which extinction curve should be used at different redshifts
or even at different luminosities. The fact that the mean
observational data have values between the intrinsic model predictions
for $\beta$ and those obtained using the SMC extinction curve, might suggest that the MW
extinction curve is not suitable for describing galaxies at $z\geq
5$. Nevertheless, the use of the SMC extinction curve appears to be
also disfavoured since model galaxies are too red compared with
observed ones. Our results suggest that the extinction curve that
might describe high redshift galaxies will have characteristics
between those of the MW and SMC ones. \citet{conroy10} found that the colours of galaxies
at $z<1$ are best described if it is assumed an attenuation curve
with a UV bump strength equal to 80 per cent of that of the MW and a
higher effective optical depth in the UV range is assumed, compared
with the MW curve.

One clear point of tension in Fig. \ref{fig:beta} is the observed
trend for $\beta$ to get bluer for fainter galaxies, something that is
not seen in the model galaxies. Nevertheless,
\citet{finkelstein12} have argued that this trend is in fact due to
assuming an average redshift for the whole sample and applying a
single correction to estimate $\beta$ as a function of the
rest-frame UV luminosity using fixed filters. \citeauthor{finkelstein12} found that once a more
consistent definition of the rest-frame UV luminosity is used, the
trend of the UV continuum slope with UV luminosity
disappears. The observations from both \citet{dunlop12} and
\citet{finkelstein12} resulted in mean
UV continuum slopes independent of the galaxy luminosity, in agreement with our
predictions. These two observational studies selected LBGs by
measuring their photometric redshifts and used the galaxy SED to obtain their absolute
rest-frame UV magnitude.

For the magnitudes measured in a particular band we do find that brighter galaxies are, on
average, more attenuated. However, the model predicts
rather grey attenuation and so, the predicted attenuations in the two
bands used to compute $\beta$ are very
similar for different HST bands, giving attenuated colours that practically do not depend
on the luminosity of a galaxy.

Fig. \ref{fig:beta} shows that extracting information about the dust
content of a galaxy based on its UV continuum slope could lead to
erroneous conclusions
if knowledge of the size distribution of dust grains and the galactic
inclination is not available. 

\subsection{The change of the UV-continuum slope with redshift.}
\begin{figure}
{\epsfxsize=8.5truecm
\epsfbox[28 5 292 281]{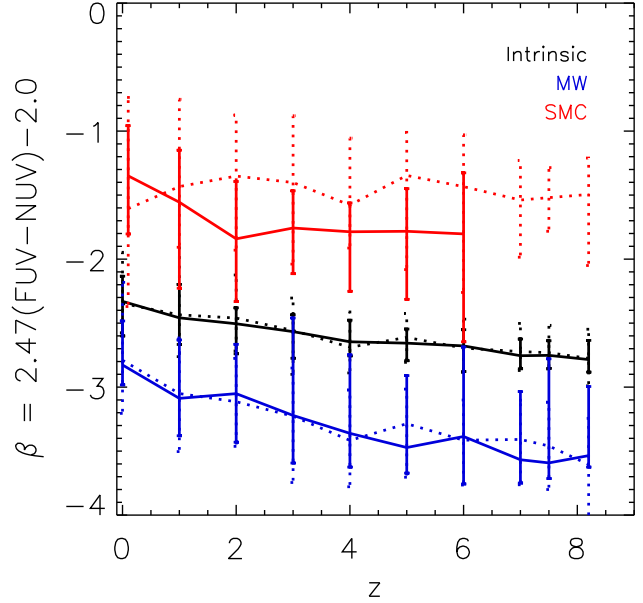}}
\caption
{The UV continuum slope of galaxies with rest-frame $M_{\rm UV}-5{\rm
  log}h\leq -20.3$ (solid lines, $L^*_{\rm UV}$ at $z=3$) and $M_{\rm UV}-5{\rm
  log}h\leq -17.8$ (dotted lines, $0.1L^*_{\rm UV}$ at $z=3$)
  measured with two rest-frame top-hat filters, NUV ($\sim 1500$\AA )
  and FUV ($\sim 2200$\AA ), as a function of redshift. The black line
  corresponds to the intrinsic UV continuum slope, the blue/red to
  that obtained using as an input for the model the MW/SMC extinction curve. 
}
\label{fig:bz}
\end{figure}

Fig. \ref{fig:bz} shows the evolution with redshift of the intrinsic UV
continuum slope, calculated with Eq.\ref{eq:b}, together with that
estimated starting from either a MW
or a SMC extinction curve for galaxies with $M_{\rm AB}$(1500\AA)$-5{\rm log}h\leq -20.3$ or $M_{\rm AB}$(1500\AA)$-5{\rm
  log}h\leq -17.8$. These two rest-frame magnitudes correspond to $L^*_{UV}$ and
$0.1L^*_{UV}$ galaxies at $z=3$ \citep{steidel99}.

The intrinsic UV continuum slope becomes bluer (i.e. $\beta$ becomes
more negative) with increasing redshift. For the intrinsic
UV continuum slope, we find a change of  0.5 magnitudes from $z=0$ to
$z=8$, independently of the absolute magnitude cut. This change is due to an increasing population of young and
massive stars at higher redshifts. This result agrees with most observational studies
\citep{bou09,finkelstein12,swuvc,bou12}, but
disagrees with the results from \citet{dunlop12} for LBGs at $5\le z
\le 7$. The fact that the intrinsic UV continuum slope for model galaxies
varies with redshift suggests that the trend found
observationally may not entirely be driven by the evolution in
the dust content of galaxies.

Fig. \ref{fig:bz} shows that the effect of using either the SMC or the
MW extinction curve impacts the predicted UV continuum slope in a
similar way to that reported previously for a fixed redshift. When
starting with the MW extinction curve we still find that the UV continuum slope becomes bluer with
increasing redshift. However, when starting with the SMC extinction curve
the variation of UV continuum slope with redshift is negligible.

\section{Conclusions}\label{sec:conclusions}

In this paper we have continued the study started by \citet{lacey11}
of Lyman-break galaxies (LBG) using {\sc galform}, the semi-analytical
model of galaxy formation originally developed by \citet{cole00}. We have
followed the galaxy formation process in the framework of the
$\Lambda$CDM cosmology. We have presented predictions for the
\citet{baugh05} variant of {\sc galform}, with the same input
parameters as those used in \citet{lacey11}. This model was
originally constructed to match the far-UV luminosity function of LBGs
only at $z=3$ but was shown to fit the observed luminosity functions
up to $z=10$, the full range currently available \citep{lacey11}.

Here we have studied the predicted rest-frame UV colours of LBGs in the range
  $2.5\le z\le 10$. The \citeauthor{baugh05} model of galaxy formation
  includes physical treatments of the hierarchical assembly of dark
  matter haloes, shock-heating and cooling of gas, star formation,
  feedback from supernova explosions, the photoionization of the
  intergalactic medium (IGM),
  galaxy mergers and chemical enrichment. The luminosities of galaxies
  are calculated from a stellar population synthesis (SPS) model and dust
  attenuation is then included using a self-consistent theoretical
  model based on the results of radiative transfer calculations. The
  far-UV dust attenuation is a critical component in any model for
  LBGs, since the effect of dust on UV light can be dramatic. 

We find that
  model galaxies have dust masses that decrease for fainter galaxies. Model galaxies with
  $-22\leq M_{AB}(1500$\AA$)-5$log$h\leq -18$ have median dust masses ranging
  from  $10^{8.5}h^{-1}M_{\odot}$ to $10^{6.5}h^{-1}M_{\odot}$ at $z=3$. We also find that median dust masses
  decrease with increasing redshift. 

We find that brighter LBGs are generally
  predicted to be more attenuated at all redshifts than fainter
  ones, though this trend is dominated by the scatter. Our results
  suggest that the LBG colour selection does not lead to an
appreciable bias in average UV attenuation of the sample. 

Our model produces galaxies with UV colours that are generally consistent with
  the colour-colour region designed observationally for selecting LBGs. From
  the model results, we expect the drop-out technique to recover most
  galaxies brighter than the absolute UV magnitude corresponding to
  the apparent magnitude limit of a particular observational survey. We find the completeness of the drop-out technique to be
  above 90 per cent at all the studied redshifts, except $z=5$, at
  which we predict a completeness above 70 per cent. These values should be considered as upper limits, since we have not
  included the effect of photometric noise, which most likely
  will push some galaxies out of the selection region at the target redshift.

We have investigated the impact that different
  parameters from the model (SPS, IGM and dust treatment) have on the UV colours, finding that
  they are most sensitive to dust and, in particular, to the
  extinction curve assumed as an input for the radiative transfer
  model with which attenuation
  curves are obtained for individual galaxies. This is related to the fact that the characteristics of
  the input extinction curves are recognisable in many of the
  attenuation curves obtained after the processing performed with the radiative transfer model from
  \citet{ferrara99}. Nevertheless, we find that the drop-out technique that
  selects LBGs based on two UV colours is very robust. 

The predicted UV luminosity function is strongly
  affected by changes in the model parameters $t_{\rm esc}$, the time
  that it takes a star to escape from the molecular cloud in which it
  is born, and
  $f_{\rm cloud}$, the fraction of gas and dust in molecular. However, we find that the predicted UV colours are not sensitive to these two parameters. Thus, although the observed UV
  colours of LBGs are reproduced by the model, the UV colours do not
  help further to constrain the dust parameters $t_{\rm esc}$ and $f_{\rm cloud}$.

We have also obtained the predicted UV continuum
slope calculated from a single rest-frame UV colour, as is commonly
done in observational studies. For galaxies at redshifts between 5 and
7 we find an intrinsic (without any attenuation) UV continuum slope of about $-2.5$, even though
some of the brightest galaxies have up to 2 magnitudes of extinction
in the UV. 
 
Using the Milky Way (MW) dust extinction curve as an input to the
radiative transfer model that calculates dust attenuation, the
predicted   UV continuum slopes are, in general, {\it bluer} than
observations. However, we find the opposite trend when the Small
Magellanic Cloud (SMC) extinction curve is used as an input. This
demonstrates the strong dependence of UV colours on dust properties
and the inadequacy of using the UV continuum slope as a tracer of dust
without further knowledge of the galaxy inclination or dust
characteristics. The mean observed UV continuum slopes are close to
the predicted intrinsic ones. Including the attenuation by dust
produces UV continuum slopes that are either too blue, if assuming a
MW extinction curve, or too red, if assuming an SMC extinction curve,
compared with the observations. This suggests that neither of these
extinction curves provides an adequate description of high redshift
galaxies. The discrepancy between the model predictions and the
observations for the slope of the attenuated UV continuum slope is
smaller than the difference between the predictions obtained using
either the MW or SMC extinction curves.

The predicted median UV continuum slope depends very weakly on the rest-frame UV
luminosity of galaxies. This result agrees with the
observational estimates from both \citet{dunlop12} and
\citet{finkelstein12}, but disagrees with other observational studies \citep{bou09,sw11,bou12}.The predicted UV continuum slope gets bluer with increasing redshift, in
agreement with many observational studies
\citep{bou09,finkelstein12,swuvc,bou12}, though this trend is not
clear when starting with the SMC extinction curve.

In conclusion, we find that the \citet{baugh05} model predicts UV
colours consistent with the 
colour-colour regions designed observationally for selecting LBGs. The
predicted intrinsic UV continuum slopes are very close to those
observed. However, once the (uncertain) dust attenuation is modelled,
the predicted and observed UV continuum slope are inconsistent. The discrepancy is
extremely sensitive to the choice of input dust extinction curve for
the radiative transfer model that calculates the dust attenuation. Actually, it is unclear which dust extinction is appropriate for high redshift galaxies. A better
knowledge of dust properties in galaxies at high redshifts will be
required to further constrain the models using LBGs.

\subsection*{ACKNOWLEDGEMENTS}
{\small
The calculations for this paper were performed on the ICC Cosmology Machine, which is part of
the DiRAC Facility jointly funded by STFC, the Large Facilities
Capital Fund of BIS, and Durham University. We thank Tom Theuns, Pratika Dayal,
Andrea Ferrara, Eli Dwek and the anonymous referee  for helpful comments. VGP acknowledges support
from the UK Space Agency. VGP, CGL and CMB acknowledge support from the Durham STFC rolling grant in theoretical cosmology.

}


\bibliographystyle{mn2e}
\bibliography{biblio}

\appendix
\section*{appendix}

Here we define explicitly the magnitude and colour cuts used to select
the model LBGs at different redshifts. The magnitude cuts have
been chosen to be either the magnitude cutoff used in the
observational studies to define their LBGs sample or the 
3$\sigma$ (or 5$\sigma$) detection limits quoted in those studies for the
bands sampling the UV continuum slope in the deepest field of the study.

\begin{itemize}
\item {\bf U-dropouts, $3\leq z\leq 3.5$, \citet{steidel95} selection:}  
  $\mathcal{R}<25.5$ and G$<27.3$ and \\
  (U$_n$-G)$>1.5$ and (U$_n$-G)$>$(G-$\mathcal{R}$)$+1.5$  and \\
  (G-$\mathcal{R}$)$<1.2$ and (G-$\mathcal{R}$)$>0$ \\

\item {\bf B-dropouts, $z\sim 4$, \citet{bou12} selection:}  
V$_{606}<30.1$ and z$_{850}<29.4$ and \\
(V$_{606}-$z$_{850}$)$<1.6$ and (B$_{435}-$V$_{606}$)$>1.1$ and \\
(B$_{435}-$V$_{606}$)$>$(V$_{606}-$z$_{850}$)$+1.1$ \\

\item {\bf V-dropouts, $z\sim 5$, \citet{bou12} selection:}  
i$_{775}<29.9$ and z$_{850}<29.4$ and \\
$[$(V$_{606}-$i$_{775}$)$>$(i$_{775}-$z$_{650}$)$+1.5$ or
(V$_{606}-$i$_{775}$)$>2$$]$ and (i$_{775}-$z$_{650}$)$<0.8$ \\

\item {\bf V-dropouts, $z\sim 5$, \citet{sw11} selection:}  
z$_{850}<27.9$ and \\
(z$_{850}-$Y$_{105}$)$<0.62$ and (V$_{606}-$i$_{775}$)$>0.81$(z$_{850}-$Y$_{105}$)$+2.14$ \\
and (i$_{775}-$z$_{850}$)$<0.76$(z$_{850}-$Y$_{105}$)$+0.83$ \\

\item {\bf i-dropouts, $z\sim 6$, \citet{bou12} selection:}  
z$_{850}<29.4$ and J$_{125}<29.9$ and\\
(i$_{775}-$z$_{850}$)$>1.3$ and (z$_{850}-$J$_{125}$)$<0.9$ \\

\item {\bf i-dropouts, $z\sim 6$, \citet{sw11} selection:}  
Y$_{105}<28.1$ and    
(Y$_{105}-$J$_{125}$)$<0.78$ and \\ (i$_{775}-$z$_{850}$)$>0.56$(Y$_{105}-$J$_{125}$)$+1.27$ \\
and (z$_{850}-$Y$_{105}$)$<0.66$(Y$_{105}-$J$_{125}$)$+0.73$ \\

\item {\bf z-dropouts, $z\sim 7$, \citet{bou12} selection:}
Y$_{105}<29.6$ and J$_{125}<29.9$ and \\  
(z$_{850}-$Y$_{105}$)$>0.7$ and (Y$_{105}-$J$_{125}$)$<0.8$ and \\
(z$_{850}-$Y$_{105}$)$>1.4$(Y$_{105}-$J$_{125}$)$+0.42$ \\

\item {\bf z-dropouts, $z\sim 7$, \citet{sw11} selection:}
J$_{125}<28.5$ and 
(J$_{125}-$H$_{160}$)$<0.95$ and \\ (Y$_{105}-$J$_{125}$)$<0.46$(J$_{125}-$H$_{160}$)$+0.89$ \\
and (z$_{850}-$Y$_{105}$)$>0.42$(J$_{125}-$H$_{160}$)$+1.48$ \\

\item {\bf Y-dropouts, $z\sim 8$, \citet{lor11} selection:}
J$_{125}<28.5$ and \\
(Y$_{105}-$J$_{125}$)$>0.9$ and (J$_{125}-$H$_{160}$)$<1.5$ and \\
(Y$_{105}-$J$_{125}$)$>0.73$(J$_{125}-$H$_{160}$)$+0.9$ \\

\item {\bf J-dropouts, $z\sim 10$, \citet{bou10} selection:}
H$_{160}<29.8$ and (J$_{125}-$H$_{160}$)$>1.2$ \\

\end{itemize}

\end{document}